\documentclass[aps,superscriptaddress,showpacs,nofootinbib,floatfix,epfs,superscriptaddress,showpacs,groupedaddress,preprintnumbers]{revtex4}
\usepackage{amsfonts,amscd,amsmath,amssymb,graphicx,color}
\usepackage[section]{placeins}
\def\et2q{\eta_{\text{\tiny QQ}}}

\def\D2q{D_{\text{\tiny QQ}}}

\def\rh{r_h}
\def\xa{x_a}
\def\rb{r_b}
\def\xb{x_b}
\def\a{\alpha}
\def\rv{r_v}

\def\se{\bar{\sigma}}
\def\sev{\bar{\sigma}_v}

\def\Td{T_{\text{\scriptsize diss}}}
\def\vl{v_{\text{\scriptsize lim}}}
\def\h0{h_0}
\def\xis{\xi_{\text{\scriptsize s}}}
\def\dm{m_{\text{\tiny D}}}

\def\oh{\frac{1}{2}}
\def\ep{\text{e}}
\def\g{\mathfrak{g}}
\def\oh{\frac{1}{2}}

\def\s{\mathfrak{s}}

\begin{document}
\title{Drag force on heavy diquarks and gauge/string duality}
\author{Oleg Andreev}
 \affiliation{L.D. Landau Institute for Theoretical Physics, Kosygina 2, 119334 Moscow, Russia}
\affiliation{Arnold Sommerfeld Center for Theoretical Physics, LMU-M\"unchen, Theresienstrasse 37, 80333 M\"unchen, Germany}
\begin{abstract} 
We use gauge/string duality to model a doubly heavy diquark in a color antitriplet moving in a thermal plasma at temperatures near the critical. With the assumption that there is no relative motion between the constituents, we calculate the drag force on the diquark. At high enough speed we find that diquark string configurations develop a cusp. In addition, we estimate the spatial string tension at non-zero baryon chemical potential, and also briefly discuss an extension to a triply heavy triquark in a color triplet. 
\pacs{11.25.Tq, 11.25.Wx, 12.38.Lg}
\preprint{LMU-ASC 07/18}
\end{abstract}
\maketitle

\section{Introduction} 
\renewcommand{\theequation}{1.\arabic{equation}}
\setcounter{equation}{0}

The concept of a diquark emerged quite naturally and inescapably as an organizing principle for hadron spectroscopy \cite{GM}.\footnote{For recent developments, see \cite{QQ-recent} and numerous references therein.} Actually, it originates from the fact that in QCD a strong force between quarks in a color antitriplet channel $\bar{\bf 3}$ is attractive. Since diquarks may be very useful degrees of freedom to focus on in QCD, it is of great importance to study them in a pure and isolated form, and determine their parameters. The latter is not straightforward because of confinement, since diquarks are colored. Of course, the same problem arises for quarks. A possible way out is to do so in the deconfined phase that allows one to study them in isolation but for temperatures above the critical (pseudocritical) temperature $T_c$. This works for quarks regardless of how high temperature is, but becomes more tricky for diquarks. The situation here is quite similar to that of $J/\Psi$ mesons. The point is that diquarks as bound states dissolve at temperatures above the dissociation temperature $\Td$. The numerical analysis shows that for bottomonium states $\Td$ is compatible to $2T_c$ \cite{qq-dis}. This opens a narrow window of opportunity for studying heavy diquarks in isolation. 

One of the profound implications of the AdS/CFT correspondence for QCD is that it resumed interest in finding its string description (string dual). This time the main efforts have been focused on a five (ten)-dimensional string theory in a curved space rather than on an old fashioned four-dimensional theory in Minkowski space. In this paper we continue these efforts. Our goal here is quite specific: to see what insight can be gained into doubly heavy diquarks moving in a hot medium by using five-dimensional effective string models. The main reasons for doing this are: (1) Because a string dual to QCD is not known at the present time. So, it is worth gaining some experience by solving any problems that can be solved within the effective string models already at our disposal. (2) Because, to our knowledge, this issue has never been discussed in the existing literature, even that related to hadron phenomenology and lattice QCD.

To contrast, what has been widely discussed is a motion of a heavy quark in a hot medium (thermal plasma) \cite{Q}. In the Langevin formalism, the motion is described by a stochastic differential equation 

\begin{equation}\label{Langevin}
\frac{d{\bf p}}{dt}=-\eta_{\text{\tiny D}}{\bf p}+\boldsymbol{\xi}(t)
\,
\end{equation}
such that the right hand side is written as the sum of a drag force, with $\eta_{\text{\tiny D}}$ a drag coefficient, and a random force. This also attracted much attention and became a hot topic in the context of AdS/CFT \cite{book}. In particular, in \cite{drag-ads} it was proposed how to calculate the drag coefficient by just considering 
\begin{figure*}[htbp]
\centering
\includegraphics[width=5.5cm]{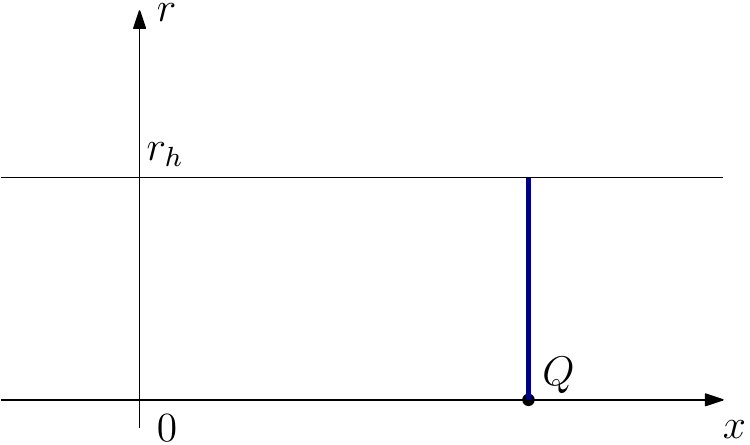}
\hspace{3cm}
\includegraphics[width=5.5cm]{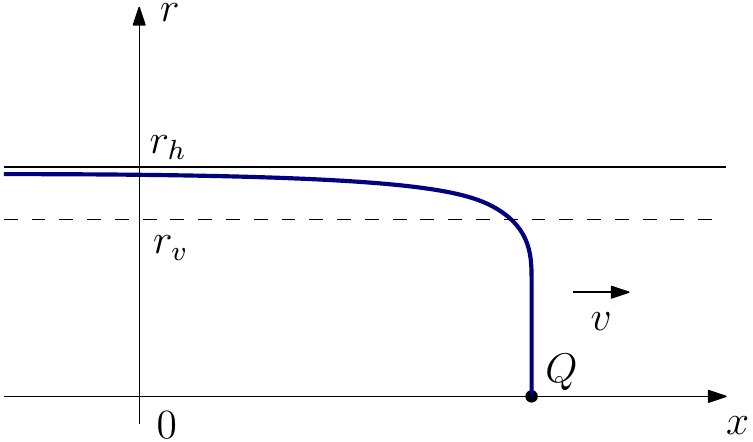}
\caption{{\small Left: A string attached to a static quark $Q$ at a point on the boundary of $\text{AdS}$ space and terminated at the black hole horizon, located at $r=\rh$ in the interior. Right: A string trailing out from a quark moving with speed $v$. The dashed line indicates 
the induced horizon related to an induced metric on the string worldsheet, as explained in the Appendix A.}}
\label{trailingQ}
\end{figure*}
a trailing string, as sketched in Figure \ref{trailingQ}. Importantly enough, for more realistic string models such a proposal gives reasonable estimates for the heavy quark diffusion coefficients which are compatible with those of lattice QCD \cite{a-D}.

The paper is organized as follows. In Section II, we begin by setting the framework and recalling some preliminary results on string configurations for diquarks. Then, we show how to calculate the drag force on a doubly heavy diquark moving in a hot medium. In Section III, we consider a concrete example to demonstrate how the proposal might work in practice. Some of the more technical details are given in the Appendices. We conclude in Section IV with a discussion of several open problems that we hope will stimulate further research.

\section{Drag force on doubly heavy diquarks as seen by string theory}
\renewcommand{\theequation}{2.\arabic{equation}}
\setcounter{equation}{0}

\subsection{Preliminaries}
For orientation, we begin by setting the framework and recalling several preliminary results. We will follow the strategy of calculating the drag force by means of a five-dimensional {\it effective} string theory that is nowadays quite common for modeling QCD at finite temperature. A key point is that the background geometry is that of a black hole. The latter implies that a dual gauge theory is in the deconfined phase.

The strings in question are Nambu-Goto strings governed by the action $S=-\tfrac{1}{2\pi\alpha'}\int d\tau d\sigma\sqrt{-\gamma}$, and living in a five-dimensional curved space with a metric of form

\begin{equation}\label{metric}
ds^2=w(r)R^2\Bigl(-f(r)dt^2+dx_i^2+f^{-1}(r)dr^2\Bigr)
\,,
\end{equation}
where $x_i=(x,y,z)$. The blackening factor $f$ is monotonically decreasing from unity to zero on the interval $[0,r_h]$. One can think of this spacetime as a deformation of the Schwarzschild black hole in $\text{AdS}_5$ space such that the boundary is at $r=0$ and the horizon at $r=\rh$. A dimensionful deformation parameter is needed to depart from conformality of $\text{AdS}_5$ for the purpose of mimicking infrared properties of QCD. One of the requirements is that the metric approaches that of the Schwarzschild black hole with $f=1-\bigl(\tfrac{r}{\rh}\bigr)^4$ as $r$ approaches the boundary. As in \cite{a-D}, we assume that this model provides a reasonable approximation to the behavior of QCD in the deconfined phase near the critical (pseudocritical) temperature. 
Importantly, the temperature of a dual gauge theory is given by the Hawking temperature of the black hole $T=\frac{1}{4\pi}\lvert \frac{d f}{d r}\rvert_{r=\rh}$. The thermal medium is assumed to be isotropic, and thus the metric is chosen to be invariant under spatial rotations. 

\begin{figure*}[htbp]
\centering
\includegraphics[width=5.75cm]{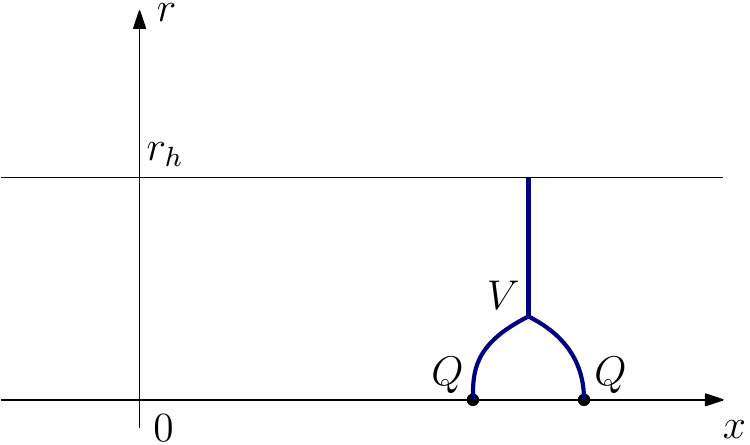}
\caption{{\small A static diquark configuration in the color $\bar{\bf 3}$ channel. The heavy quarks $Q$ are placed on the boundary of space. The two strings attached to the quarks and one attached to the horizon join at the baryon vertex $V$ in the interior.}}
\label{QQv0}
\end{figure*}

In contrast to the single quark case, this is not the whole story. The reason is that the stringy construction of diquark states in color antitriplet demands a new object \cite{a-screen}. It is a baryon vertex (string junction), at which three strings meet together. In five dimensions, the resulting construction is sketched in Figure \ref{QQv0}. Noteworthy, it 
leads to a result for the diquark free energy which is consistent with lattice simulations \cite{a-screen}.

From a string theory point of view, such a vertex is a five-brane wrapped on a five-dimensional internal manifold in ten-dimensional space \cite{witten-bv}. In the above construction we assume that quarks are placed at the same point in the internal space and, therefore, the detailed structure of this space is not important for us, except a possible warp factor depending on the radial coordinate $r$. Note that the vertex can be regarded as point like in five-dimensional space. 

Taking the leading term in the $\alpha'$-expansion of the world-volume action for the brane and choosing static gauge, results in an effective action for the baryon vertex

\begin{equation}\label{vertex}
S_{\text{vert}}=-m\int d t\sqrt{f}\,{\cal V}(r)
\,.
\end{equation}
Here $m$ is a parameter which takes account of a volume of the internal space together with possible $\alpha'$-corrections in the brane action.\footnote{On a way to the string dual to QCD, $\text{AdS}_5\times\mathbf{S}^5$ is supposed to be replaced by its deformation such that $\mathbf{S}^5\rightarrow\mathbf{X}$, with $\mathbf{X}$ a new compact space.} The form of ${\cal V}$ is determined by the warp factor of the metric. It is also worth bearing in mind that $S_{\text{vert}}$ is a function of the radial coordinate $r$.

\subsection{Calculating the drag force}

\begin{figure*}[htbp]
\centering
\includegraphics[width=6.25cm]{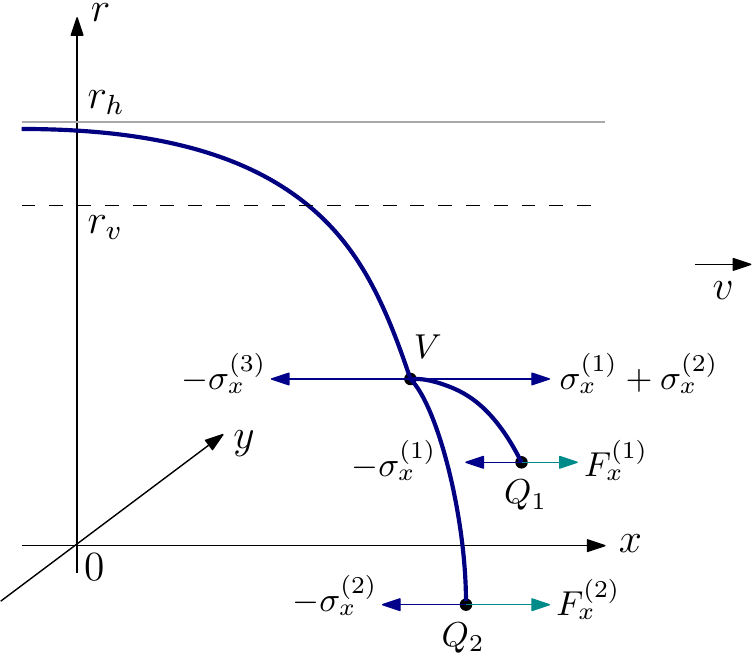}
\caption{{\small A diquark configuration moving with speed $v$ in the $x$-direction. The heavy quarks are set on the boundary. $V$ is a baryon vertex located at $r=r_0$ in the interior. It is connected to the quarks by the two strings. The remaining third string is trailing out from the vertex. External forces $F_x^{(i)}$ exerted on the quarks are needed to balance internal string tensions $\sigma_i$.}}
\label{QQ}
\end{figure*}

We now want, as in Figure \ref{trailingQ} for a single quark, to consider a diquark configuration moving with speed $v$. In doing so, we can without loss of generality place heavy quark sources at 
arbitrary points on the $xy$-plane, at $r=0$, and consider their uniform motion with speed $v$ in the $x$-direction, as sketched in Figure \ref{QQ}.
We will work in the lab frame related to the thermal medium of a dual gauge theory and assume that there is no relative motion between the quarks.

Let us begin by explaining, in a qualitative way, what one might expect from such a configuration. The argument goes as follows. Since the motion is uniform, nothing happens with the shape of the two strings connected to the quarks. In fact, what the external forces do is that they pull the remaining (trailing) string in the $x$-direction, and this costs energy. Then, the same energy-based argument that one uses in the single quark case shows that in both cases the drag forces are equal in magnitude. This argument is robust to the variations of string shapes (for the strings terminated on the quarks) and of the form of ${\cal V}$.

It is natural to think of the above argument as a starting point for understanding the diquark case. To understand that better, let us translate it into mathematical language. First, for each string we take the static gauge\footnote{If the context is clear, we sometimes omit the index labelling the strings.}

\begin{equation}\label{gauge}
t(\tau)=\tau\,,\qquad
r(\sigma)=a\sigma+b
\,,
\end{equation} 
where $(\tau,\sigma)$ are worldsheet coordinates. Combining these with the boundary conditions

\begin{equation}\label{boundary-c}
r_1(0)=r_2(0)=0\,,\qquad r_1(1)=r_2(1)=r_3(0)=r_0\,,
\qquad r_3(1)=\rh
\,,
\end{equation}
we get $a_1=a_2=b_3=r_0$, $a_3=\rh-r_0$, and $b_1=b_2=0$. Then, for each string we choose the ansatz 
\begin{equation}\label{ansatz0}
x(r)=vt+\xi(r)
\,,\qquad
y(r)=\eta(r)
\,,
\end{equation}
which is a slightly extended version of that originally proposed in \cite{drag-ads}. 

In the presence of external forces exerted on the quarks, the total action ${\cal S}$ is the sum of three Nambu-Goto actions, an effective action of the baryon vertex, and boundary terms arising from the coupling with external fields

\begin{equation}\label{action}
{\cal S}=\sum_{i=1}^3  S_i +S_{\text{\tiny vert}}+S_{\text{\tiny b}}
\,.
\end{equation}
Taking the metric \eqref{metric} and using the ansatz \eqref{ansatz0}, we find that each Nambu-Goto action takes the form

\begin{equation}\label{action-t}
S=-\g\int dt\int dr \,w \sqrt{1-\frac{v^2}{f}+f(\partial_r\xi)^2+\bigl(f-v^2\bigr)(\partial_r\eta)^2}
\,\,,
\end{equation}
with $\g=\frac{R^2}{2\pi\alpha'}$. Similarly, for the baryon vertex, taking a five-brane effective action along with the ansatz $x=vt$, we find that the action \eqref{vertex} now becomes 

\begin{equation}\label{vertex-v}
S_{\text{vert}}=-m\int d t\sqrt{f-v^2}\,{\cal V}(r_0)
\,.
\end{equation}

Actually, what we have written above, though incomplete, is enough for our purposes because we do not need the explicit form of $S_{\text{\tiny b}}$. We just need the fact that the variation of ${\cal S}$ is given by\footnote{The sign conventions for the $\sigma$'s and $F$'s are those of Figure \ref{QQ}.}

\begin{equation}\label{variation-t}
\delta{\cal S}=-\sum_{i=1}^3 \int dt \int dr\, 
\bigl(\partial_r\sigma^{(i)}_x \,\delta\xi_i+\partial_r\sigma^{(i)}_y \,\delta\eta_i\bigr)
+
\int dt\, \bigl(F_x^{(1)}-\sigma_x^{(1)}\bigr)\delta x_1+\bigl(F_x^{(2)}-\sigma_x^{(2)}\bigr)\delta x_2
+
\bigl(\sigma_x^{(1)}+\sigma_x^{(2)}-\sigma_x^{(3)}\bigr)\delta x_0
+\dots
\,\,,
\end{equation}
where the dots stand for other boundary contributions. This formula is derived at time $t=0$ when the $x$-coordinates of the quarks and vertex were $x_i$ and $x_0$, respectively. Since the motion is uniform, all the equations obtained via the variational principle are time independent. Looking at the right hand side of equation \eqref{variation-t}, we can immediately obtain the first integrals of the equations of motion: $\sigma_x=c$ and $\sigma_y=c'$, where $c$ and $c'$ are constants and 

\begin{equation}\label{sigma-i}
\sigma_x=-\frac{\g w f\partial_r\xi}{\sqrt{1-\frac{v^2}{f}+f(\partial_r\xi)^2+\bigl(f-v^2\bigr)(\partial_r\eta)^2}}
\,,
\qquad
\sigma_y=-\frac{\g w \bigl(f-v^2\bigr)\partial_r\eta}{\sqrt{1-\frac{v^2}{f}+f(\partial_r\xi)^2+\bigl(f-v^2\bigr)(\partial_r\eta)^2}}
\,.
 \end{equation}
It is easy to write these formulas in another form, more convenient for our analysis

\begin{equation}\label{deri}
(\partial_r\xi)^2=\frac{1}{f}\frac{1-\frac{v^2}{f}}{\frac{\se^2}{\sigma_x^2}-1-(1-\frac{v^2}{f})^{-1}\frac{\sigma_y^2}{\sigma_x^2}}
\,,\qquad
(\partial_r\eta)^2=\frac{\sigma_y^2}{\sigma_x^2}
\frac{(\partial_r\xi)^2}{\bigl(1-\frac{v^2}{f}\bigr)^2}
\,,
\end{equation}
where $\se=\g w\sqrt{f}$ is an effective string tension \cite{az3}.

Now let us consider the trailing string. It is helpful to first analyze the case $c'=0$.
From eqn.\eqref{sigma-i}, it follows that $\eta$ is constant. This means that the string lies on the $xr$-plane. If so, then we can proceed along the lines of \cite{drag-ads}, briefly reviewed in the Appendix A. The key point is that the only way to have $(\partial_r\xi)^2> 0$ is to set

\begin{equation}\label{string3}
\sigma_x^{(3)}=\sev
\,,
\end{equation}
where $\sev=\se (\rv)$ and $\rv$ is a solution of the equation $f=v^2$. Such a solution always exists because $f$ monotonically decreases from $1$ to $0$ on the interval $[0,\rh]$. 

Now, taking $c'\not=0$, the function $1-\frac{v^2}{f}$ is positive on the interval $[0,\rv]$. The function $\frac{\se^2}{\sigma_x^2}-1-(1-\frac{v^2}{f})^{-1}\frac{\sigma_y^2}{\sigma_x^2}$ goes to plus infinity as $r$ goes to zero and to minus infinity as $r$ goes to $\rv$.\footnote{Note that the effective string tension $\se$ for  $r\rightarrow 0$ behaves as $\se\sim 1/r^2$.} From this, it follows that it has one or more zeros on $[0,\rv]$. If so, then it is negative on 
the interval between the largest zero and endpoint $r=\rv$. As a result, $(\partial_r\xi)^2<0$ on such an interval, and, therefore there is no trailing solution.

The conditions for mechanical equilibrium can be deduced from the vanishing of the boundary terms in the variation of the total action. In particular, eqn.\eqref{variation-t} implies that 

\begin{equation}\label{fb0}
F_x^{(1)}-\sigma_x^{(1)}=0
\,,\qquad 
F_x^{(2)}-\sigma_x^{(2)}=0
\,
\end{equation}
for the quarks, and 

\begin{equation}\label{fbv}
\sigma_x^{(1)}+\sigma_x^{(2)}-\sigma_x^{(3)}=0
\,
\end{equation}
for the vertex. From these formulas, it follows that the total force exerted on the diquark 
is $F_{\text{\tiny tot}}=F_x^{(1)}+F_x^{(2)}=\sigma_x^{(3)}$, with $\sigma_x^{(3)}$ given by \eqref{string3}. Using the fact that $F_{\text{\tiny drag}}=-F_{\text{\tiny tot}}$, we get

\begin{equation}\label{drag}
F_{\text{\tiny drag}}=-\sev
\,.
\end{equation}
We have obtained precisely the result, as anticipated before. It can be written more explicitly as $F_{\text{\tiny drag}}=-\g w(\rv)v$. The same expression was also found in \cite{drag-kir} for a single quark. We have thus shown that if there is no relative motion between the quarks, then the drag force on a diquark is equal to that on a single quark.

\section{A Concrete Example}
\renewcommand{\theequation}{3.\arabic{equation}}
\setcounter{equation}{0}
 In Section II, we obtained the formula for the drag force acting on a doubly heavy diquark without paying much attention to the problem of existence of the corresponding string configuration. Now we will to some extent settle the problem using a simple example. Once we do this, we find a rather non-trivial structure of the diquark configuration even in that case. In particular, a cusp develops on one of the strings at high enough speed.
\subsection{The model}
For the warp and blackening factors, we set 

\begin{equation}\label{fS}
w(r)=\frac{\ep^{\s r^2}}{r^2}\,,
\qquad
f(r)=1-\Bigl(\frac{r}{\rh}\Bigr)^4
\,,
\end{equation}
which can be obtained from the Euclidean metric of \cite{az2} by the standard Wick rotation $t\rightarrow it$. One might think of criticizing such a choice on the grounds of the consistency of the string sigma model at the loop level, but there is not any restriction at the classical level where we are working. The advantages are two-fold. First, it allows a great deal of simplification of the resulting equations, that makes it useful for understanding more complicated forms of $w$ and $f$. Second, it provides quite plausible results for $SU(3)$ pure gauge theory which are, in some cases, amazingly close to those of lattice gauge theory \cite{pureYM}.

Given the explicit form of $f$, it is straightforward to determine the Hawking temperature as a function of $\rh$, and also find the position of the induced horizon 

\begin{equation}\label{rs}
T=\frac{1}{\pi\rh}
\,,
\qquad
\rv=\rh\sqrt[4]{1-v^2}
\,.
\end{equation}
We also take note of the fact that with the above choice for the warp and blackening factors, the critical temperature is given by $T_c=\frac{\sqrt{\s}}{\pi}$ \cite{az2}. From this formula and \eqref{rs}, it is found convenient to introduce a dimensionless parameter $h=\s\rh^2=\frac{T^2_c}{T^2}$. So, $h<1$ in the deconfined phase.

We complete the model by specifying the gravitational potential for the vertex. Following \cite{3q}, we take it to be of the form

\begin{equation}\label{vertex-example}
{\cal V}(r)=\frac{\ep^{-2\s r^2}}{r}
\,.
\end{equation}
Apart from the brane construction, another crucial reason for this is the ability of the model to describe the lattice data for the three-quark potentials \cite{3q}.

\subsection{A diquark configuration moving along the quark-quark axis}

As an example we consider a diquark moving at a constant speed along its axis. Our aim is two-fold, first to address the problem of existence of the diquark configuration, and secondly to see what kind of length contraction one could expect in the model under consideration. Of course, this can not be the whole story because of the diquark ability to move in an arbitrary direction with respect to the axis. The latter greatly adds to the complexity of the problem but would not affect our main conclusions, as we expect based on the results for a quark-antiquark pair in which the direction of the hot wind has only a little effect \cite{book}. 

\begin{figure*}[htbp]
\centering
\includegraphics[width=4.9cm]{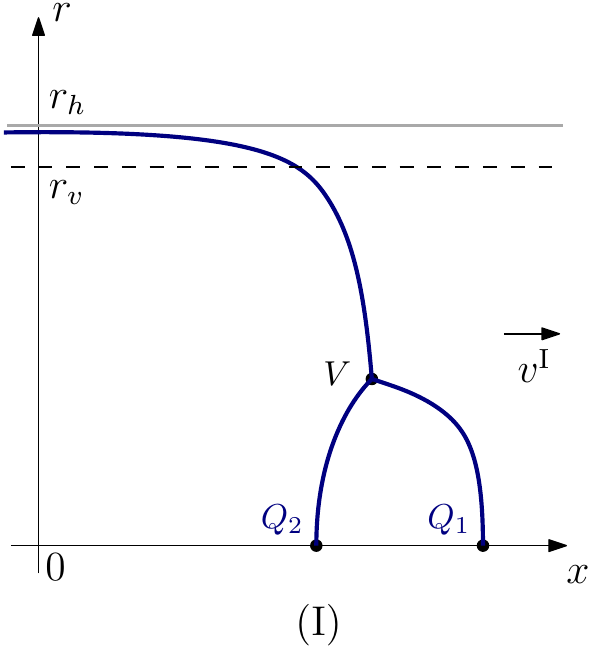}
\hspace{1cm}
\includegraphics[width=4.9cm]{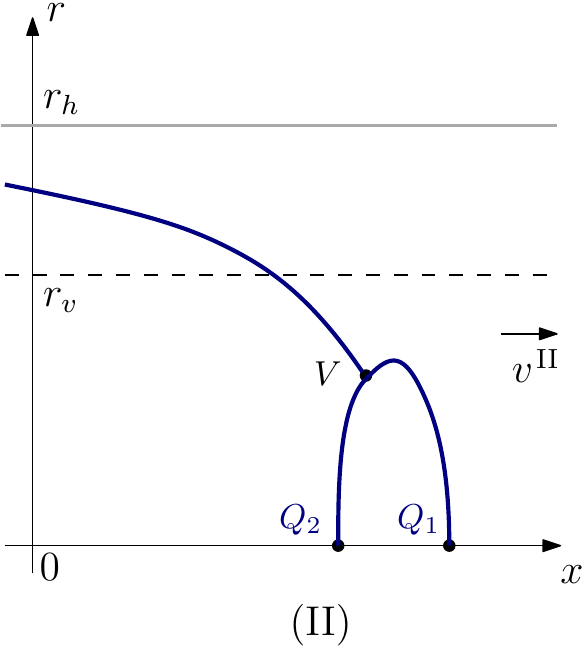}
\hspace{1cm}
\includegraphics[width=4.9cm]{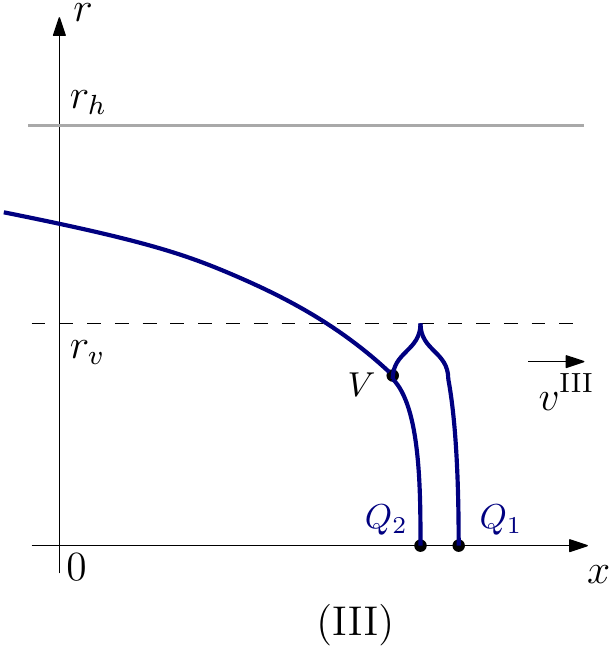}
\caption{{\small A planar diquark configuration moving in the $x$-direction. Here the $r$-coordinate of the vertex is fixed at $r_0$, while $v$ takes different values such that $v^{\text{I}}<v^{\text{II}}<v^{\text{III}}$. In the case III, a cusp forms at a point lying on the induced horizon.}}
\label{conI-III}
\end{figure*}

We take, without loss of generality, the diquark axis to be the $x$-axis. Since the speed of the diquark is directed in the $x$-direction, on symmetry grounds the corresponding string configuration is planar. It lies on the $xr$-plane, as sketched in Figure \ref{conI-III}. In this Figure we do
not indicate the external forces and internal string tensions explicitly. Like in the previous section, we work in the lab frame and assume that the quark relative speed is zero.

\subsubsection{Configuration I}

Consider the static configuration sketched in Figure \ref{QQv0}. To get further, we need to boost it along the $x$-axis with a small speed $v$. In this case we can get a good idea of what to expect based on the experience gained from the trailing string picture shown in Figure \ref{trailingQ}. Thus it seems reasonable that very little happens with the strings terminated on the quarks, while the string previously terminated on the horizon drastically changes its shape and becomes a trailing string. At late times the resulting configuration will be like that shown in Figure \ref{conI-III} on the left.

Since the configuration moves uniformly, we can invoke the formula \eqref{l12} to write the distance between the quarks as 

\begin{equation}\label{lI}
\ell(\nu;v,h)=\sqrt{\frac{h}{\s}}
\int^{\nu}_0 
\frac{d\rho}{1-\rho^4}
\sqrt{1-v^2-\rho^4}
\Biggl[
\biggl(\frac{\nu^4(1-\rho^4)}{\rho^4(1-\nu^4)}\frac{\ep^{2h(\rho^2-\nu^2)}}{\sin^2\theta_1}
-1\biggr)^{-\oh}
+
\biggl(\frac{\nu^4(1-\rho^4)}{\rho^4(1-\nu^4)}\frac{\ep^{2h(\rho^2-\nu^2)}}{\sin^2\theta_2}-1\biggr)^{-\oh}
\Biggr]
\,,
\end{equation}
where $\nu=\frac{r_0}{\rh}$ and $\theta_i$ is an angular parameter corresponding to the $i$-string. Clearly, $\ell$ is a function of several variables.

The parameters $\theta_i$ are, however, subject to the constraints which are the gluing conditions \eqref{fb-x} and \eqref{fb-r}. In the model we are considering, these conditions take the form

\begin{subequations}
\begin{gather}
\sin\theta_1+\sin\theta_2+\phi=0
\,,\label{fxI}
\\
\cos\theta_1+\cos\theta_2-\sqrt{1-\phi^2}-3\kappa\sqrt{1-\nu^4}
\biggl(1+4h\nu^2+\frac{2\nu^4}{1-v^2-\nu^4}\biggr)\ep^{-3h\nu^2}=0
\,,\label{frI}
\end{gather}
\end{subequations}
with 
\begin{equation}\label{theta3}
\phi=\frac{v\nu^2\,\ep^{h(\sqrt{1-v^2}-\nu^2)}}{\sqrt{(1-v^{2})(1-\nu^{4})}}
\,.
\end{equation}
With this at hand, $\ell$ can be viewed as a function of three variables $\nu$, $v$ and $h$.

The important observation for what follows is that the first string changes its shape with increase of speed. The transition from configuration I to configuration II occurs at a certain "critical" speed between slow and medium and has a clear geometrical interpretation. It happens when $\tan\alpha_1=\infty$ or, equivalently, $\cos\theta_1=0$.\footnote{The convention adopted in the Appendix A is that $\theta_1$ is negative. So, this corresponds to $\sin\theta_1=-1$.}  Loosely speaking, the endpoint $V$ looks like a turning point of a function $r(\xi)$, as approached from the right. Therefore there is no contribution from string 1 stretched between $Q_1$ and $V$ to the force balance along the $r$-direction. A more formal way to say this is to write eqs.\eqref{fxI} and \eqref{frI} as 

\begin{equation}\label{v1}
\sqrt{\phi(2-\phi)}
-
\sqrt{1-\phi^2}
-3\kappa\sqrt{1-\nu^4}
\biggl(1+4h\nu^2+\frac{2\nu^4}{1-v^2-\nu^4}\biggr)\ep^{-3h\nu^2}
=0
\,.
\end{equation}
A solution $v(\nu,h)$ of this equation is a "critical" speed above which string 1 is bent into a U-shape.

\subsubsection{Configuration II}

Similarly, for configuration II, we use eqs. \eqref{l12} and \eqref{l4} to deduce the formula for the distance between the quarks

\begin{equation}\label{lII}
\begin{split}
\ell(\nu;v,h)
=&
\sqrt{\frac{h}{\s}}
\int^{\nu}_0 
\frac{d\rho}{1-\rho^4}
\sqrt{1-v^2-\rho^4}
\Biggl[
\biggl(\frac{\nu^4(1-\rho^4)}{\rho^4(1-\nu^4)}\frac{\ep^{2h(\rho^2-\nu^2)}}{\sin^2\theta_1}
-1\biggr)^{-\oh}
+
\biggl(\frac{\nu^4(1-\rho^4)}{\rho^4(1-\nu^4)}\frac{\ep^{2h(\rho^2-\nu^2)}}{\sin^2\theta_2}-1\biggr)^{-\oh}
\Biggr]\\
+&
2\sqrt{\frac{h}{\s}}\int^{\lambda}_\nu\frac{d\rho}{1-\rho^4}
\sqrt{1-v^2-\rho^4}
\biggl(\frac{\nu^4(1-\rho^4)}{\rho^4(1-\nu^4)}\frac{\ep^{2h(\rho^2-\nu^2)}}{\sin^2\theta_1}
-1\biggr)^{-\oh}
\,.
\end{split}
\end{equation}
Here $\lambda=\frac{r_c}{\rh}$, with $r_c$ a coordinate of the turning point (see Figure \ref{conf4}). Importantly, $r_c$ obeys the condition \eqref{rm} which now can be written as

\begin{equation}\label{lambda}
\bigl(1-\lambda^{-4}\bigr)\ep^{2h\lambda^2}=\sin^2\theta_1\bigl(1-\nu^{-4}\bigr)\ep^{2h\nu^2}
\,.
\end{equation}
It allows one to reduce the number of parameters describing the configuration by one.

We can further reduce the number of parameters by the gluing conditions. One of those is written above as \eqref{fxI}, and the other is obtained from \eqref{fb-r2}

\begin{equation}\label{frII}
\cos\theta_1-\cos\theta_2+\sqrt{1-\phi^2}+
3\kappa\sqrt{1-\nu^4}\biggl(1+4h\nu^2+\frac{2\nu^4}{1-v^2-\nu^4}\biggr)\ep^{-3h\nu^2}
=0
\,.
\end{equation}
As in the previous subsection, it is possible, at least numerically, to solve the system of gluing equations to get a reduced description with fewer parameters. Thus we can write $\ell$ as a function of threes variables $\nu$, $v$, and $h$.

As speed increases, string 2 stretched between $Q_2$ and $V$ becomes more and more straight, until at angle $\theta_2=0$ the transition to configuration III occurs at a certain "critical" speed between medium and high. Setting $\theta_2=0$ in eqs.\eqref{fxI} and \eqref{frII}, we find  

\begin{equation}\label{v2}
1-2\sqrt{1-\phi^2}
-
3\kappa\sqrt{1-\nu^4}
\biggl(1+4h\nu^2+\frac{2\nu^4}{1-v^2-\nu^4}\biggr)\ep^{-3h\nu^2}
=0
\,.
\end{equation}
A solution $v(\nu,h)$ is a "critical" speed above which the diquark configuration looks like that of Figure \ref{conI-III} on the right. 

It is noteworthy that at the transition from configuration II to configuration III string 1 touches the induced horizon. This follows from eqs. \eqref{fxI} and \eqref{lambda}. Indeed, a short calculation shows that $\lambda=\sqrt[4]{1-v^2}$. If so, then as explained in the Appendix A, a cusp forms exactly at the touching point, and the corresponding cusp angle is 

\begin{equation}\label{cuspII}
\tan\gamma=\frac{1}{v}\biggl(\frac{1-v^2}{1-2hv^2\sqrt{1-v^2}}\biggr)^{\oh}
\,.
\end{equation}
It is well-defined because $h<1$.

\subsubsection{Configuration III}

Now consider configuration III. While the two previous configurations do not involve something so unusual, this does. The first point is that a cusp appears on string 1. In contrast to the 
case of motion with the "critical speed" given by \eqref{v2}, for higher speed the cusp angle is simply

\begin{equation}\label{c-angle}
\gamma=0
\,,
\end{equation}
as it follows from \eqref{cusp4c}.

To obtain from the formulas \eqref{l12} and \eqref{l4-1} the expression for the distance between the quarks, we will have to subtract one expression from another, as clear from Figure \ref{conI-III} on the right. As a result, we find

\begin{equation}\label{lIII}
\begin{split}
\ell(\nu;v,h)
=&
\sqrt{\frac{h}{\s}}
\int^{\nu}_0 
\frac{d\rho}{1-\rho^4}
\sqrt{1-v^2-\rho^4}
\Biggl[
\biggl(\frac{\nu^4(1-\rho^4)}{\rho^4(1-\nu^4)}\frac{\ep^{2h(\rho^2-\nu^2)}}{\sin^2\theta_1}
-1\biggr)^{-\oh}
-
\biggl(\frac{\nu^4(1-\rho^4)}{\rho^4(1-\nu^4)}\frac{\ep^{2h(\rho^2-\nu^2)}}{\sin^2\theta_2}-1\biggr)^{-\oh}
\Biggr]\\
+&
2\sqrt{\frac{h}{\s}}\int^{\sqrt[4]{1-v^2}}_\nu\frac{d\rho}{1-\rho^4}
\sqrt{1-v^2-\rho^4}
\biggl(\frac{\nu^4(1-\rho^4)}{\rho^4(1-\nu^4)}\frac{\ep^{2h(\rho^2-\nu^2)}}{\sin^2\theta_1}
-1\biggr)^{-\oh}
\,.
\end{split}
\end{equation}

The second point is that there is a limiting speed, above which $\ell$ vanishes

\begin{equation}\label{limv}
\ell(\nu;\vl,h)=0
\,.
\end{equation}
What happens is that as speed increases, quark $2$ "approaches" closer and closer to quark $1$ and finally "collides" with it. This intuitively obvious fact can be checked numerically using eqs. \eqref{lIII} and \eqref{limv}. The results show that $\vl$ may be well below unity.\footnote{See also Figure \ref{pdiag} on the left.} From this, one can draw the conclusion that the length contraction is not identical with that of special relativity. Of course, it is expected that the length contraction in the thermal medium need not be Lorentzian type since the medium defines a preferred rest frame.

We conclude our discussion of configuration III with a remark on what needs to be said about the gluing conditions at the vertex. In fact, these conditions are similar to those of configuration II and, therefore, given by eqs. \eqref{fxI} and \eqref{frII}.

\subsubsection{Some analysis}

Let us first recall the description in \cite{a-screen} of the static diquark configuration of Figure \ref{QQv0}. In that analysis, we showed that in the $\bar{\bf 3}$ channel there are two branches of $\ell(\nu;h)$ and each of those is bounded from above. This means that the connected configuration exists only if $\ell$ does not exceed the critical distance. We pick up the branch of the solution which leads to $F_{\text{\tiny QQ}}^{\bar{\bf 3}}\sim-\frac{1-\dm\ell}{\ell}$ as $\ell\rightarrow 0$, with $\dm$ the Debye mass. In this case, the corresponding formulas can be obtained from those of subsection 1 by setting $v=0$. The parameter $\nu$ takes values in the interval 
$[0,\nu_\ast]$. The function $\ell (\nu;h)$ grows with $\nu$ so that $\ell(\nu_\ast;h)$ is the critical distance at given temperature. Note that $\nu_\ast$ is determined by temperature.

In addition, let us make a simple estimate of the dissociation temperature based on $\xis(\Td)\sim 2\varrho$, where $\xis$ is a screening length and $\varrho$ a diquark radius at zero temperature. Treating the screening length as the Debye length $\xis=1/\dm$ and using the expression for the Debye mass \cite{a-screen}

\begin{equation}\label{mD}
\frac{\dm}{T}=a-b\frac{T_c^2}{T^2}
\,,
\end{equation}
we get 
\begin{equation}\label{Tdis}
\Td\sim\frac{1}{4a\varrho}\Bigl(1+\sqrt{1+16ab\varrho^2T_c^2}\Bigr)
\,.
\end{equation}
For $SU(3)$ pure gauge theory with $T_c=260\,\text{MeV}$, $a=0.71$ and $b=0.24$. This yields 
$\Td\sim 2.1\,T_c$ ($cc$) and $\Td\sim 2.9\,T_c$ ($bb$), with the radii $0.28\,\text{fm}$ ($cc$) and $0.19\,\text{fm}$ ($bb$) obtained from an effective quark-quark potential of the "Cornell" Coulomb+linear form at half strength \cite{QQ-recent}. As expected, the bottomonium diquarks dissociate at higher temperatures. These simple estimates are quite similar to those based on the analysis of the Schr\"odinger equation for diquark states in two flavor QCD \cite{qq-dis}.

\begin{figure*}[htbp]
\includegraphics[width=5.8cm]{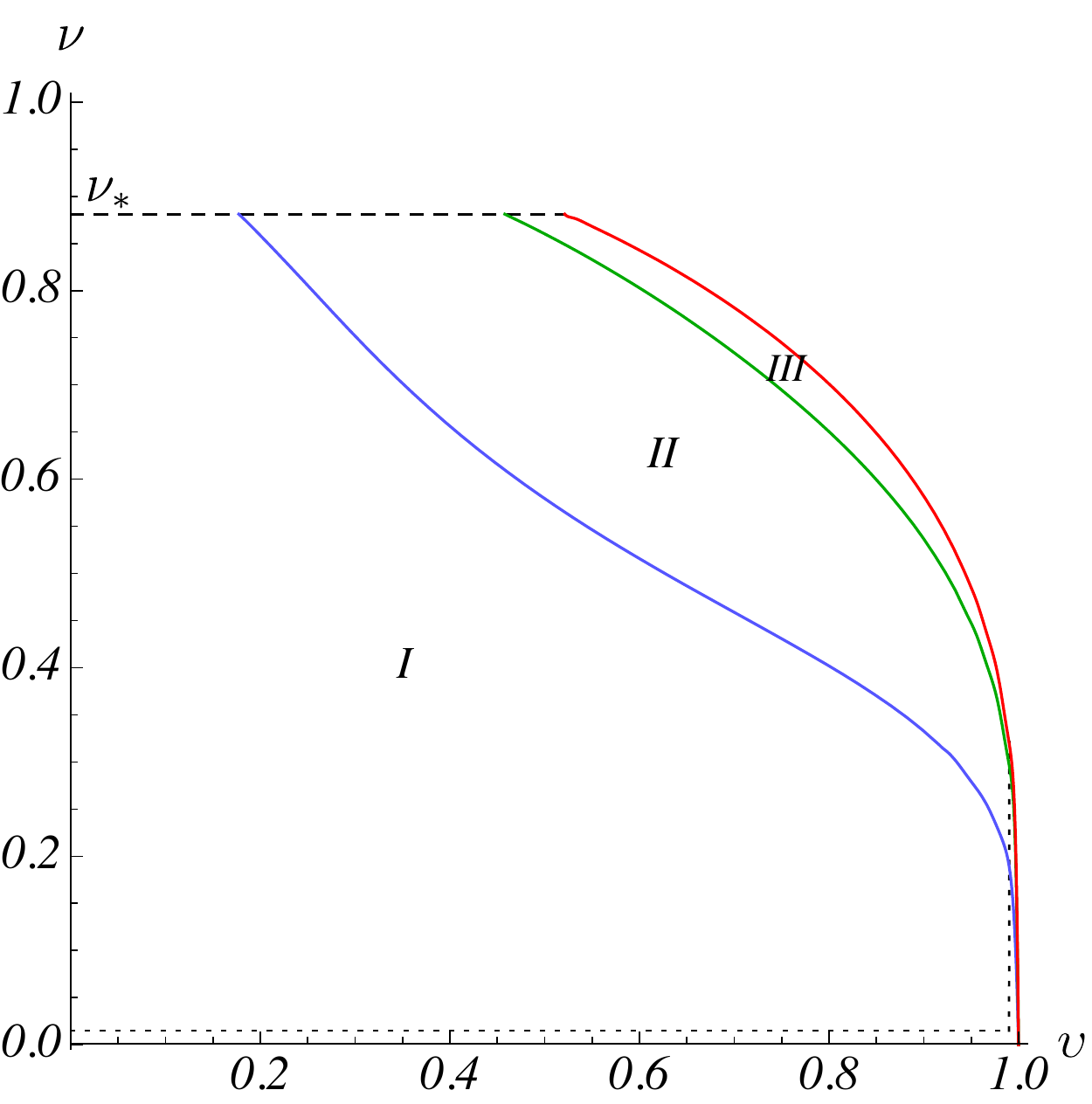}
\hspace{3cm}
\includegraphics[width=5.7cm]{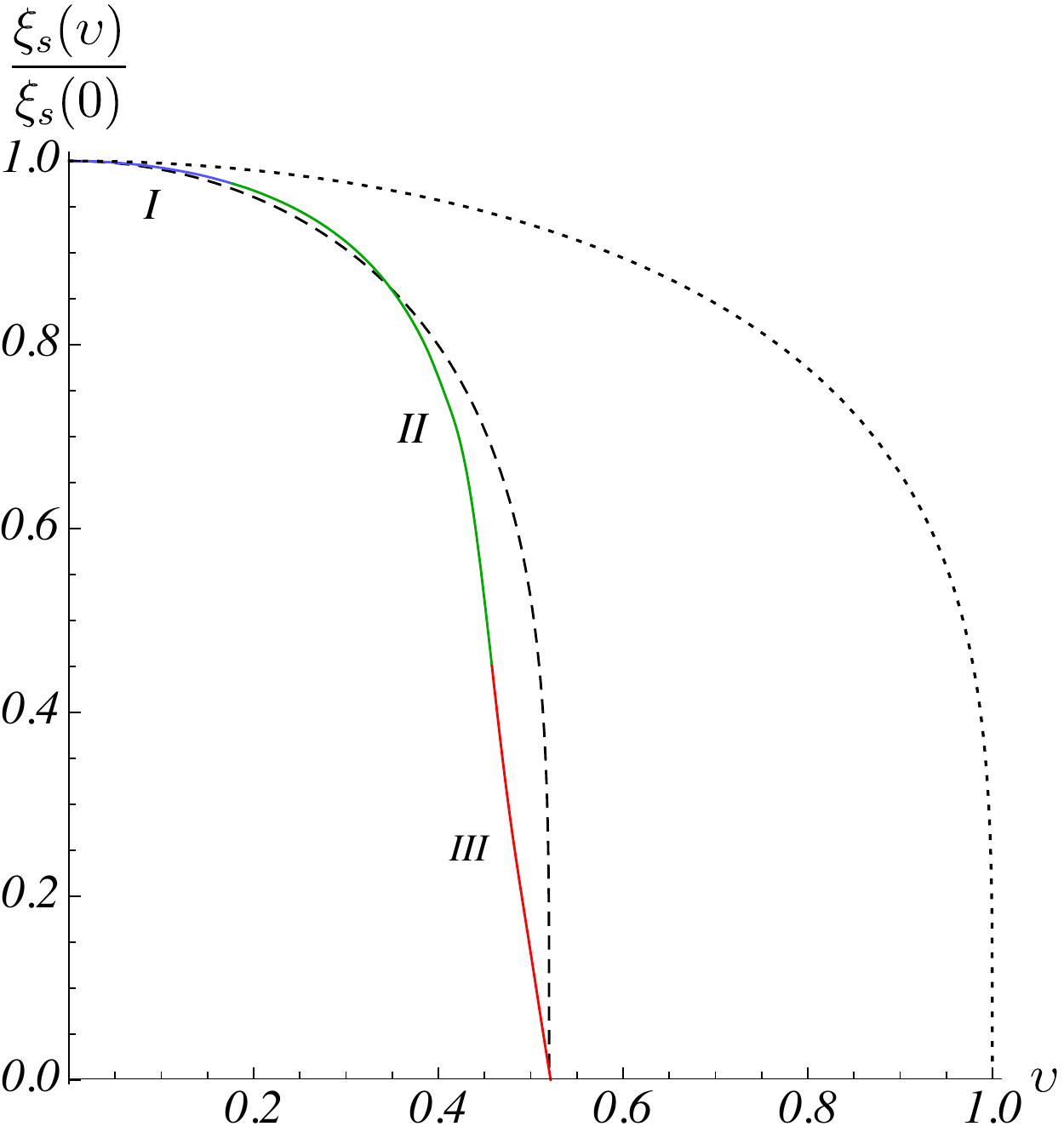}
\caption{{\small At $h=\frac{4}{9}$, and thus $\nu_\ast\approx 0.88$. The labeling is according to Figure \ref{conI-III}. Left: A temperature slice of the parametric space. The dashed line represents the maximal value of $\nu$. The dotted lines indicate cutoffs. The colored curves from left to right correspond to \eqref{v1}, \eqref{v2}, and \eqref{limv}. Right: $\xis=\ell(\nu_\ast;v,h)$ as a function of $v$. Here $\vl\approx 0.52$. Each of the different colored parts of the curve is plotted by using the corresponding formula for $\ell$. The dotted and dashed curves are defined by \eqref{uaw} and \eqref{di-uaw}.}}
\label{pdiag}
\end{figure*}

Now suppose that the static configuration is boosted so that it moves along the quark-quark axis. In this situation (at late times) there are the three possible shapes of the configuration, as shown in Figure \ref{conI-III}. Each can be described by the three parameters $\nu$, $v$, and $h$. In general, the parameter space of the model is quite complicated, but there exist some relatively simple divisions. For example, when the temperature range is narrow and restricted to a region above $T_c$, say $T_c\lesssim T\lesssim 3T_c$, there are no singularities.\footnote{Note that it corresponds to $\frac{1}{9}\lesssim h\lesssim 1$.} This range is of great interest, as the dual gauge theory is so strongly coupled that it makes its study difficult. In Figure \ref{pdiag} on the left, we 
present a two-dimensional slice of the parametric space which contains three regions according 
to a number of the string configurations of subsections $1-3$. It is actually nothing special, but a couple of points are worthy of note.

From the gluing conditions \eqref{frI} and \eqref{frII}, it follows that there are no solutions with $\nu=\sqrt[4]{1-v^2}$. Such a curve always lies outside the allowed region of 
the parametric space and, therefore, is not shown in the Figure. In other words, the baryon vertex never touches the induced horizon. It is always located between the induced horizon and boundary. This implies that the right hand side of \eqref{vertex-v} is real and, therefore, the tachyonic instability does not occur.

The effective string model we are considering here suffers from power divergences and requires a suitable regularization. These divergences are usually associated with infinitely heavy quark sources placed on the boundary. One way to deal with this situation is to impose a lower bound (cutoff) $\epsilon$ on the radial coordinate $r$. If so, then $\nu$ is also bounded from below by $\frac{\epsilon}{\rh}$. The corresponding cutoff is indicated by the dotted horizontal line. But this is not the whole story for the diquark configuration. The point is that the induced horizon approaches the boundary as speed increases. Clearly, the string construction make sense only when the induced horizon is above the cutoff, namely $\rv>\epsilon$. This leads to a upper bound on $v$. In particular, with \eqref{rs}, it is given by $v<\sqrt{1-\frac{\epsilon^4}{\rh^4}}$. This is indicated by the vertical dotted line. Since the calculation of the drag force does not suffer from the problem of infinities, there is no need for us to go further into this aspect.

The function $\ell(\nu;v,h)$ is rather complicated and defined piecewise by the corresponding integral expression in each region of the parameter space. We will now analyze it for large and small $\nu$. Both are of special interest for phenomenology.

We begin with large $\nu$. The alternative to the Debye screening length defined directly from the correlator of Polyakov loops is to define the screening length as $\xis\equiv\ell(\nu_\ast;v,h)$ \cite{wind}.\footnote{Even in the AdS case this definition requires a caveat. The problem is incompleteness at larger $\nu$. One way to address this problem is to introduce one more diagram as suggested in \cite{yaffe}. However, it remains to be seen whether both definitions of the screening length will agree.} For a quark-antiquark pair, such a definition yields a peculiar form of length contraction 

\begin{equation}\label{uaw}
\xis(v)\approx\sqrt[4]{1-v^2}\,\xis(0)
\,.
\end{equation}
In Figure \ref{pdiag} on the right, we plot $\xis$ given piecewise by the formulas \eqref{lI}, \eqref{lII}, and \eqref{lIII} at $\nu=\nu_\ast$. Obviously, it is much different from what is expected from \eqref{uaw}. The reason is simple and straightforward: the existence of a limiting speed. It thus makes sense to take this fact into account and consider a modification of \eqref{uaw}

\begin{equation}\label{di-uaw}
\xis(v)\approx\sqrt[4]{1-\frac{v^2}{\vl^2}}\,\xis(0)
\,.
\end{equation}
As seen from the Figure, this is a quite acceptable approximation, especially when speed is slower than $0.45$. 

\begin{figure*}[htbp]
\centering
\includegraphics[width=9.75cm]{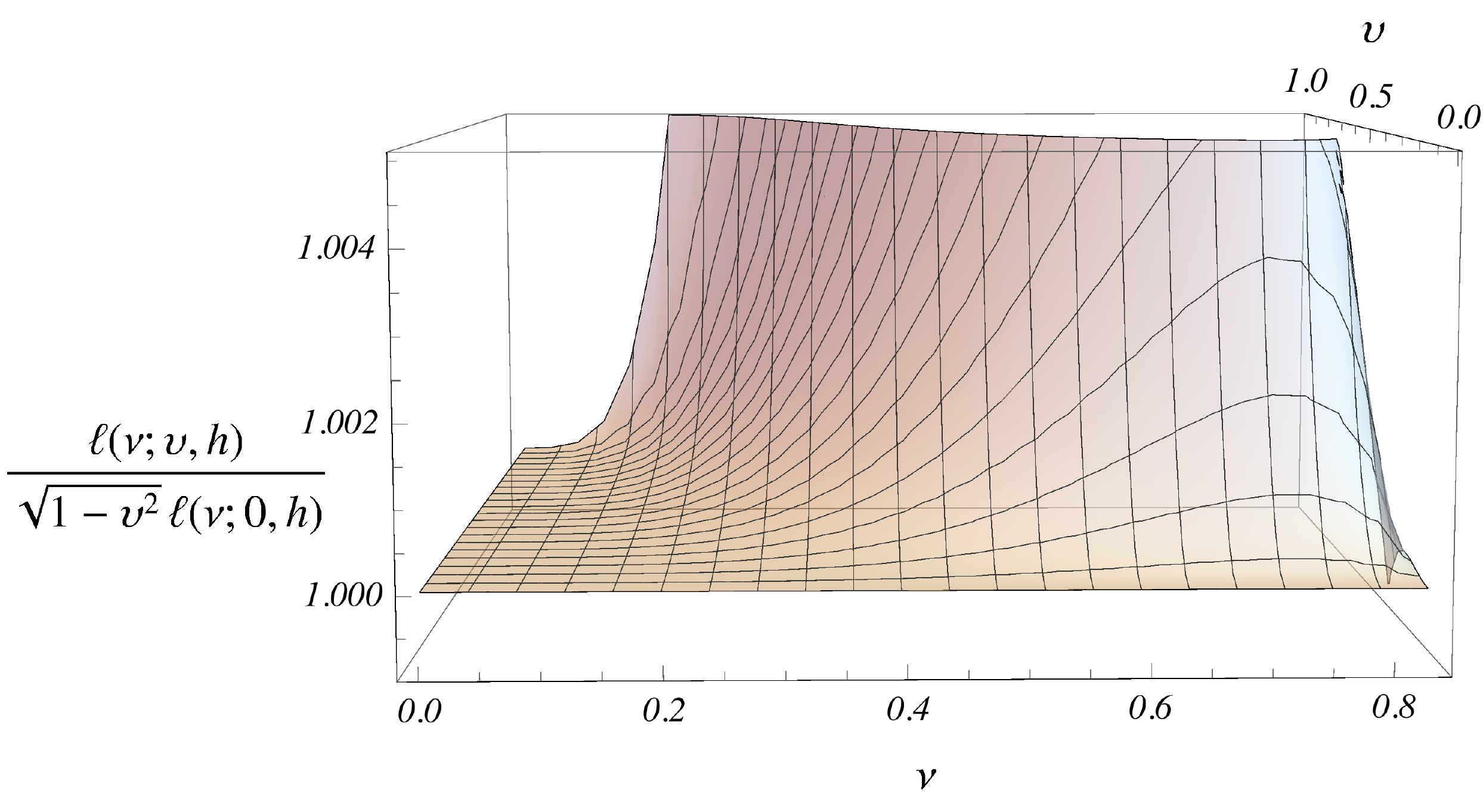}
\caption{{\small $\ell(\nu;v,h)/\sqrt{1-v^2}\ell(\nu;0,h)$ as a function of $\nu$ and $v$ at $h=\frac{4}{9}$.   }}
\label{lor}
\end{figure*}

Next, let us consider what happens when $\nu$ is small. It follows from the structure of the parametric space, and in particular from the left panel of Figure \ref{pdiag}, that in this 
case only region I matters, so that $\vl\rightarrow 1$ as $\nu\rightarrow 0$. The analysis of eq.\eqref{lI} for small $\nu$ is simple, and the first two Taylor coefficients are computed in the Appendix C. So one is led to suspect that in this limit 

\begin{equation}\label{Lorentz}
\ell(\nu;v,h)\approx\sqrt{1-v^2}\,\ell(\nu;0,h)
\,,
\end{equation}
that is nothing else but the Lorentz contraction formula. In Figure \ref{lor}, we plot the ratio 
$\ell(\nu;v,h)/\sqrt{1-v^2}\ell(\nu;0,h)$ at $h=\frac{4}{9}$ which enables us to gain more detailed insight into the approximation \eqref{Lorentz}. The plot is restricted to the range $1$ to $1.005$. We see that the $v$ dependence is in fact of Lorentzian form up to $\nu=0.15$, but for larger values it gets more involved. Heuristically, we can say that the formula \eqref{Lorentz} holds as long as the vertex remains much closer to the boundary than to the black hole horizon.

Now a question arises: What approximation might be relevant to the Debye screening length, as considered in \cite{a-screen}? No real answer will be given here but we plan to address this issue in future research.

\section{Concluding Comments}
\renewcommand{\theequation}{4.\arabic{equation}}
\setcounter{equation}{0}

(i) The purpose of this paper has been to initiate discussions on heavy diquarks as probes of strongly coupled plasma. As the first step in this direction, we considered the drag force on a test diquark by using the five-dimensional effective string models. Certainly, there are many topics and issues which  deserved to be studied in depth and with the greatest seriousness. Some of those are mentioned in \cite{book}, although in the context of heavy quarks, in particular momentum broadening and disturbance of the plasma induced by test probes. 

(ii) The non-relativistic limit of \eqref{drag} is that the frag force is proportional to the spatial string tension.\footnote{In the AdS case, this is always true. No matter what speed a single quark (diquark) is moving \cite{sin}.} We have \cite{a-D}

\begin{equation}\label{drag-c}
F_{\text{\tiny drag}}=-\sigma_s v
\,,
\end{equation}
with $\sigma_s=\g w(\rh)$ the spatial string tension. This follows from the fact that $\rv\rightarrow\rh$ as $v\rightarrow 0$.

Thus, when the motion is non-relativistic, the spatial string tension plays a pivotal role 
in controlling how strong the drag force is. Using the models at our disposal, it is possible to make some estimates. In \cite{a-D}, we got a couple of estimates at zero baryon chemical potential $\mu$. The goal here is to make a simple estimate of $\sigma_s$ at non-zero chemical potential. To this end, we follow the lines of \cite{a-screen} and involve the one-parameter deformation of the Reissner-Nordstr\"om-AdS black hole solution \cite{chamblin}, with the warp factor $w(r)=\frac{\ep^{\s r^2}}{r^2}$, where $\s$ is a deformation parameter.\footnote{In \cite{a-screen}, this model was used to estimate the Debye screening masses near the critical line in the $\mu T$-plane of two flavor QCD \cite{dor-mu}. Now we wish to supplement our list with the estimate of the spatial string tension.} As explained in the Appendix C of \cite{a-screen}, one can think of $T$ and $\mu$ as functions of two parameters 

\begin{equation}\label{tm}
T(h,q)=\frac{9}{8\pi}\frac{\sqrt{\s}(1-q^2)\,h^{\frac{3}{2}}}{\ep^{\frac{3}{2}h}-1-\frac{3}{2}h}
\,,\qquad
\mu(h,q)=\frac{2\sqrt{3\s}\mathfrak{r}\,q \bigl(1-\ep^{-\frac{1}{2}h}\bigr)}{\bigl(9+(7+6h)\ep^{-2h}-16\ep^{-\frac{1}{2}h}\bigr)^{\frac{1}{2}}}
\,.
\end{equation}
Here, as before, $h=\s\rh^2$, while $q$ and $\mathfrak{r}$ are new parameters. The parameter $q$ is  related to a black hole charge. It takes values in the interval $[0,1]$. The parameter $\mathfrak{r}$ is considered as free. 

Given the formal formulas that we have described above, it seems straightforward to apply those and determine the spatial string tension as a function of $T$ and $\mu$. In the left panel of Figure 
\begin{figure*}[htbp]
\centering
\includegraphics[width=7cm]{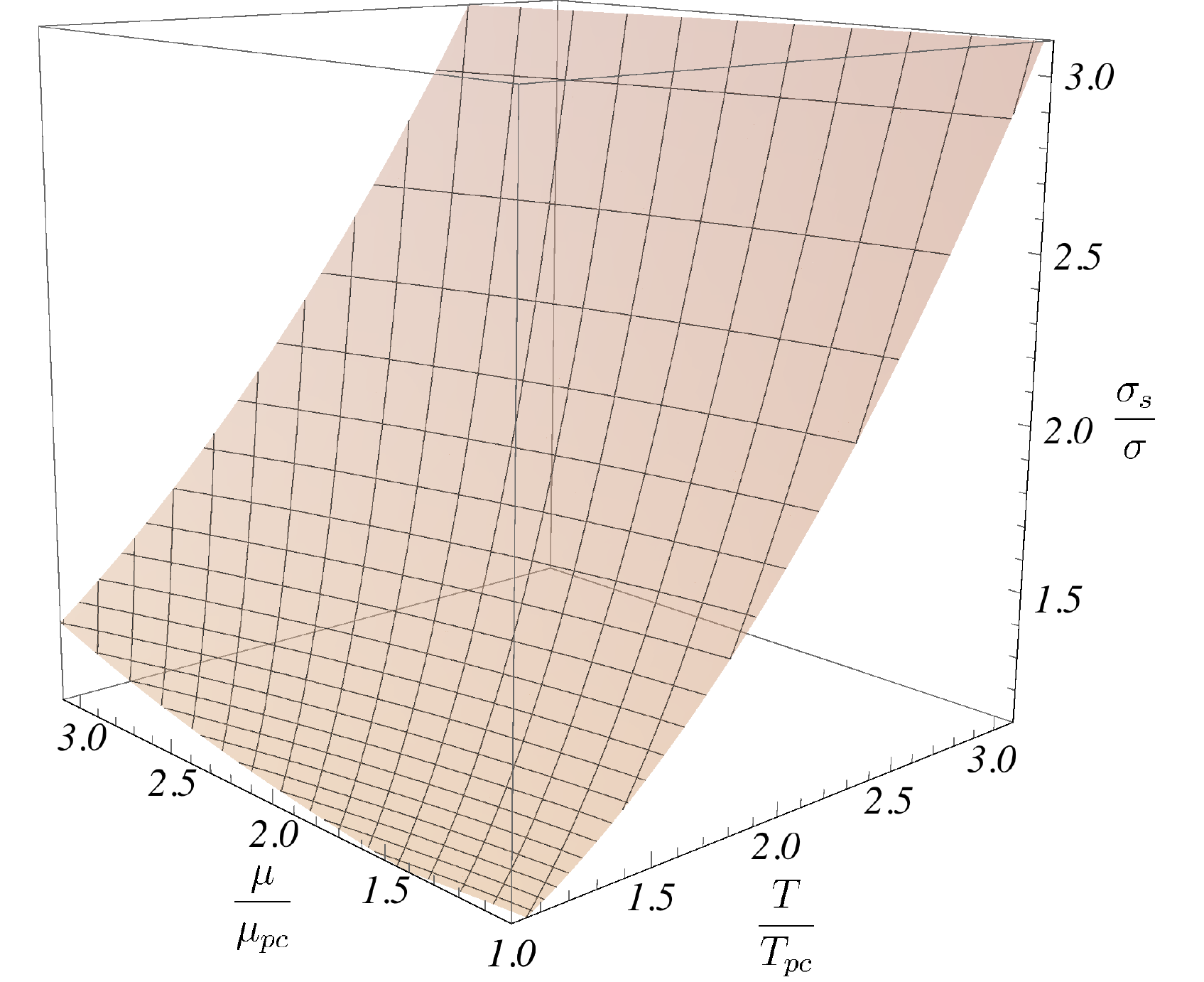}
\hspace{2.25cm}
\includegraphics[width=6.75cm]{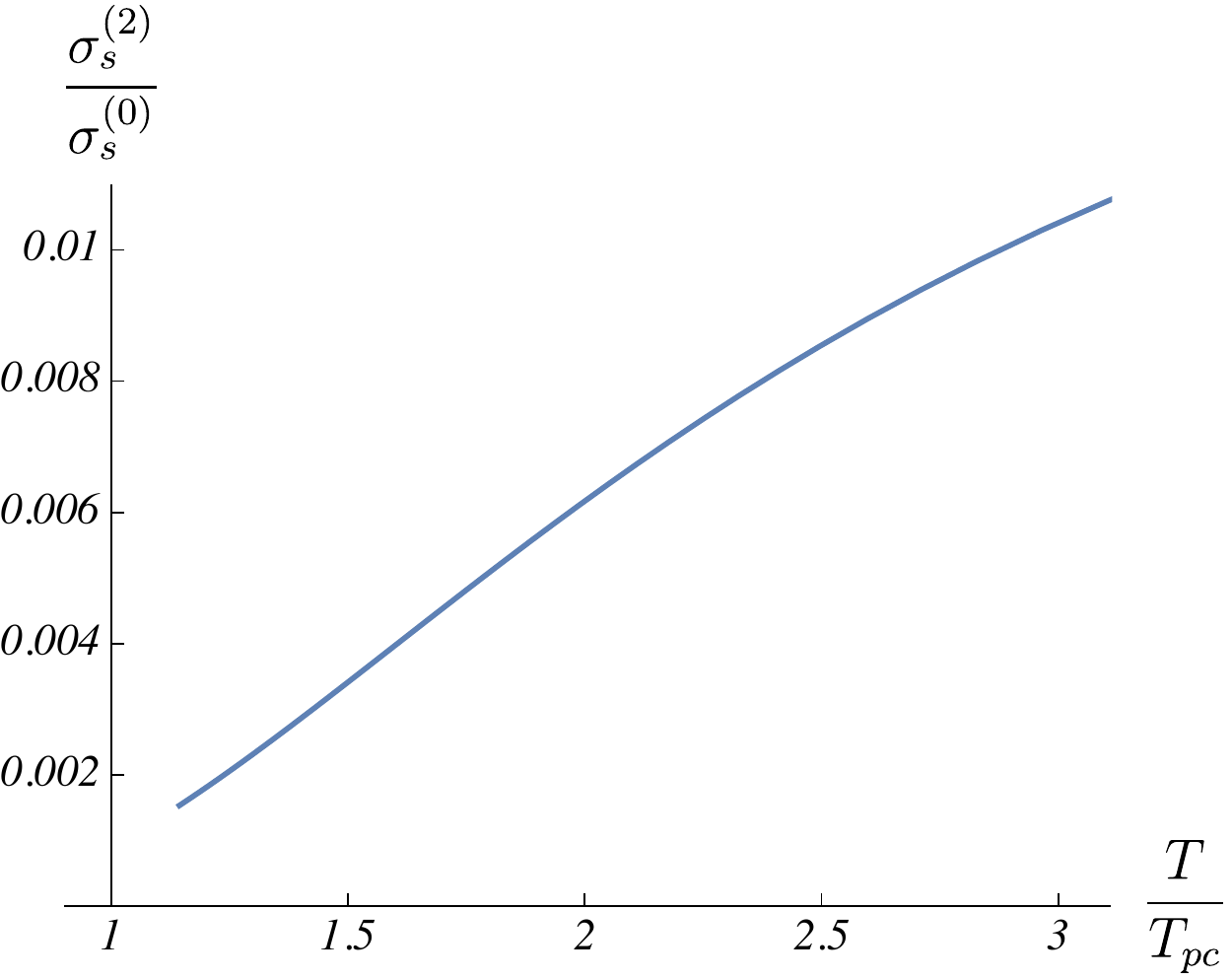}
\caption{{\small Left: Ratio $\frac{\sigma_s}{\sigma}$ versus temperature and baryon chemical potential. Here $\mathfrak{r}=6$. Right: Ratio $\frac{\sigma_s^{(2)}}{\sigma_s^{(0)}}$ versus temperature.}}
\label{s2/s0}
\end{figure*}
\ref{s2/s0} 
we plot this function, normalized by the string tension $\sigma$ at zero temperature and chemical potential.\footnote{Such a tension is simply $\sigma=\g\s\ep$.} Units are set by $T_{pc}=0.26\sqrt{\s}$ and $\mu_{pc}=2\sqrt{\s}$, as in \cite{a-screen} for two flavor QCD \cite{dor-mu}. We see that $\sigma_s$ is increasing with $T$, and more slowly with $\mu$. Additionally, we consider the Taylor expansion of $\sigma_s(T,\mu)$ about the point $(T,0)$. In the model under consideration, it is given by 

\begin{equation}\label{sigma-mu}
\sigma_s(T,\mu)=\sum_{n=0}^\infty \sigma_s^{(2n)}\Bigl(\frac{\mu}{T}\Bigr)^{2n}
\,.
\end{equation}
The coefficients $\sigma_s^{(2n)}$ can directly be derived from \eqref{drag-c} and \eqref{tm}. The first two are presented in the Appendix D. Note that in this expansion odd powers of $\mu$ do not appear because $h$ is an even function of $\mu$. Since small $\mu$ is of great interest for lattice QCD, in the right panel of Figure \ref{s2/s0} we present the result for the ratio $\frac{\sigma_s^{(2)}}{\sigma_s^{(0)}}$. 

Both results are predictions, as so far there are no lattice studies available. However, this requires a caveat: in contrast to lattice QCD, the model we are considering has no explicit dependence on quark masses. 

(iii) Another interesting concept is that of a triquark. It naturally occurs in hadron spectroscopy, for example, by treating pentaquark states as bound states of color antitriplet diquarks and color triplet triquarks.\footnote{For some examples, see \cite{triquark}.} 

\begin{figure*}[htbp]
\centering
\includegraphics[width=5.85cm]{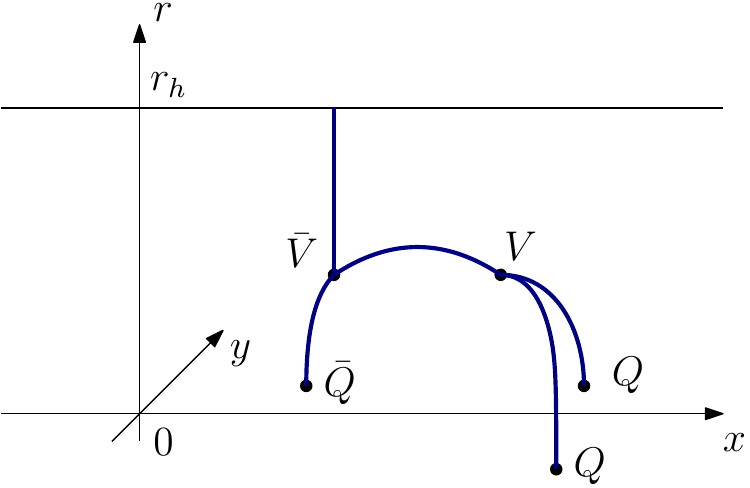}
\hspace{2.75cm}
\includegraphics[width=5.85cm]{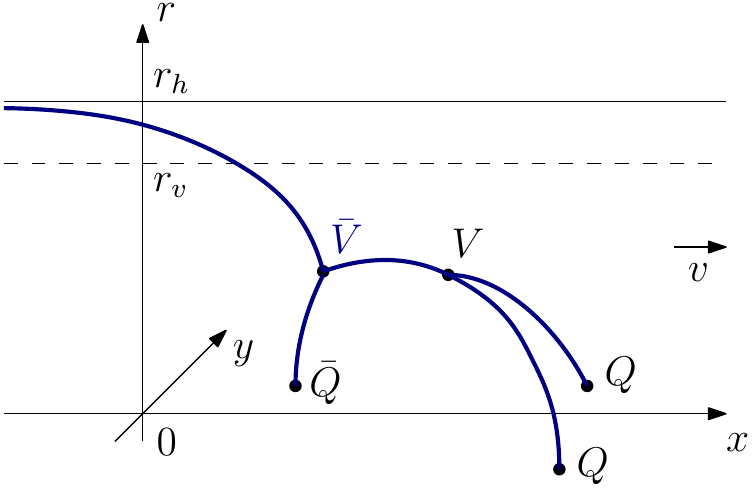}
\caption{{\small String configurations of heavy triquarks in the color $\bf 3$ channel. The heavy quarks $Q$ and antiquark $\bar Q$ are placed on the boundary of space. The strings are stretched between the quark sources, vertices and horizon. Left: A static configuration. Right: A configuration moving with speed $v$ in the $x$-direction.}}
\label{trailing3Q}
\end{figure*}

In the static case, it is straightforward to build a string configuration for a triply heavy triquark in the deconfined phase, as one shown in the left panel of Figure \ref{trailing3Q}. What distinguishes it from 
the previous configurations is that it contains an antibaryon vertex $\bar V$, that is, it is a vertex connecting three antiquarks to form an antibaryon. As in Section III, we should now ask what is the drag force acting on a triquark moving in a hot medium. To this end, we consider a triquark configuration moving with speed $v$, as sketched in the right panel of Figure \ref{trailing3Q}. If we proceed in this way, the arguments similar to those of Section III show that the drag force is given by \eqref{drag}. In other words, if the quark's relative motion is negligible, then the drag force is independent of a number of constituents. What, however, matters is a number of trailing strings (representation of color group). In all the cases, there is a single trailing string that corresponds to the color triplet/antitriplet channel. 

It is worthy of noting that there is one aspect to a triquark string configuration which is peculiar to the case of any multiquark configuration and that is its instability due to $V\bar V$ annihilation. One example of this is a well-known flip-flop for the tetraquark potential. This has a natural interpretation in string theory \cite{a-baryons}. The point is that $V\bar V$ is nothing else but a brane-antibrane system. For such a system there is a critical separation such that a tachyonic mode develops for smaller separations. The tachyonic instability represents a flow toward annihilation of the brane-antibrane system. If  it occurs in a triquark configuration, the configuration decays to two configurations, one for a single quark and one for a meson. We will not pursue this circle of ideas any further here and leave them for future study.

(iv) By now there is reasonably strong evidence, mainly from lattice QCD, that the SU(3) theory of quarks and gluons has a dual description in terms of quantum strings. Since the precise formulation of the latter is still not known, what we can do is to gain useful insight and grow with each success of the effective string model already at our disposal. One might think of criticizing this model on several grounds, in particular because of the use of string theory to describe a gauge theory with a finite number of colors. This is in fact related to the long-standing question, namely whether $N_c = 3$ is good enough for the $1/N_c$ expansion in QCD. We have nothing to say at this point, except that we hope to return to this issue in future work.

\begin{acknowledgments}
We would like to thank I. Aref'eva, P. de Forcrand, S. Hofmann, R. Metsaev, P. Weisz, and U.A. Wiedemann for useful and encouraging discussions. We also thank the Arnold Sommerfeld Center for Theoretical Physics and CERN Theory Division for the warm hospitality. This work was supported in part by Russian Science Foundation grant 16-12-10151.
\end{acknowledgments}

\appendix
\section{A Nambu-Goto string moving with constant speed}
\renewcommand{\theequation}{A.\arabic{equation}}
\setcounter{equation}{0}
Here we explore in some detail a few examples of strings steadily moving in the five-dimensional spacetime whose metric is given by \eqref{metric}. This provides the basis for the practical calculations of Section III. We consider the case, where the horizon is closer to the boundary than the so-called soft wall. This guarantees that a dual gauge theory is deconfined \cite{az2}.

The motion of strings is studied using the Nambu-Goto action

\begin{equation}\label{NG}
S=-\frac{1}{2\pi\alpha'}\int_0^1 d\sigma\int d\tau\,\sqrt{-\gamma}
\,,\qquad
\gamma=\det\gamma_{\alpha\beta}
\,,
\end{equation}
where $\alpha'$ is a string constant, $(\tau,\sigma)$ are worldsheet coordinates, and $\gamma_{\alpha\beta}$ is an induced metric.

\subsubsection*{1. Single string configurations}

\begin{figure*}[htbp]
\centering
\includegraphics[width=4.95cm]{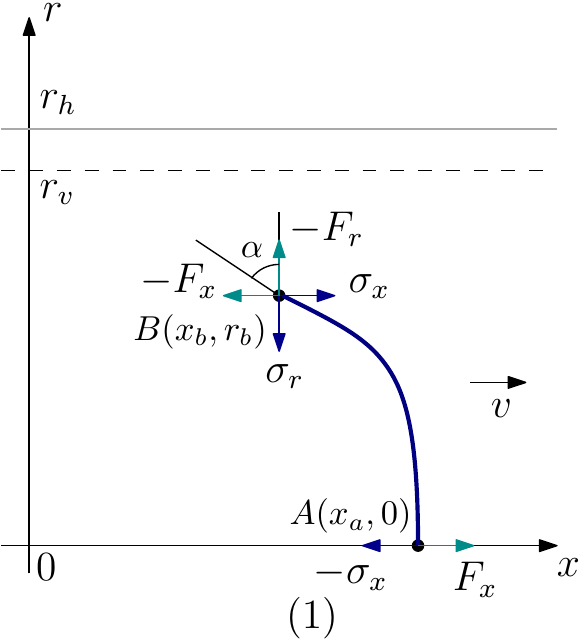}
\hspace{1.2cm}
\includegraphics[width=4.95cm]{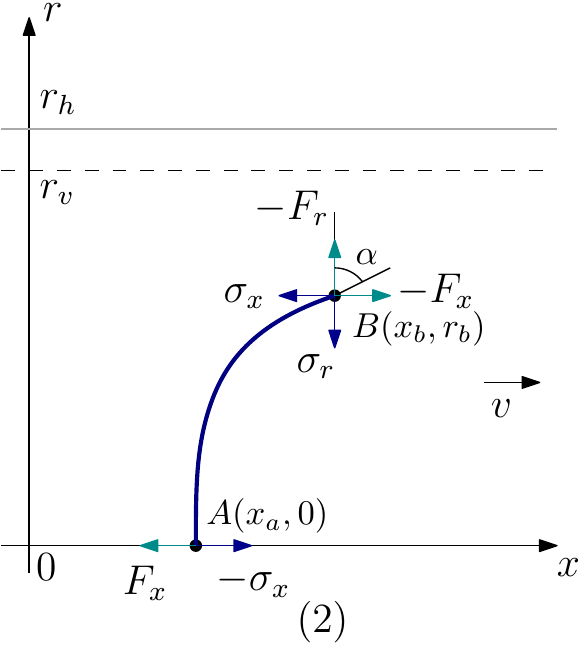}
\hspace{1.2cm}
\includegraphics[width=4.95cm]{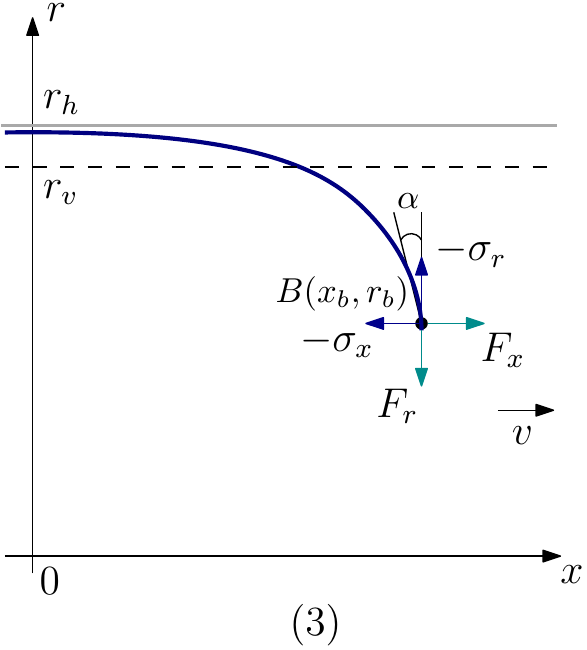}
\caption{{\small Single string configurations at time $t=0$. In all the cases the string moves with speed $v$ in the $x$-direction. $\rh$ and $\rv$ are the horizon locations of the spacetime and induced metrics, respectively. $F_i$ stand for some external forces exerted on the string endpoints to balance internal string tensions $\sigma_i$ and keep the string in uniform motion. In configurations $1$ and $2$ the string is of finite length and is stretched between the points $A$ and $B$. In configuration 3 the string is of infinite length and is trailing out behind the endpoint $B$. }}
\label{conf1-3}
\end{figure*}

We begin with what are perhaps the elementary examples. These are the cases, denoted in Figure \ref{conf1-3} as configurations $1$ and $2$, 
in which one string endpoint is on the boundary at $r=0$, while another in the bulk at $r=r_b$. In each case, the string lies on the $xr$-plane and uniformly moves in the $x$-direction with speed $v$. We consider a lab frame which is at rest and related to the thermal medium of a dual gauge theory. 

Since we are interested in uniform motion, we choose the static gauge $t=\tau$ and $r(\sigma)=c_1\sigma+c_0$. Combining the latter with the boundary conditions at the endpoints 

\begin{equation}\label{bcond}
r(0)=0\,,
\qquad r(1)=r_b
\,,
\end{equation}
we obtain $c_1=r_b$ and $c_0=0$. For $x(t,r)$, we take the ansatz of \cite{drag-ads} 

\begin{equation}\label{ansatz}
x(t,r)=vt+\xi(r)
\,.
\end{equation}
In the light of the applications of string theory to heavy ion collisions, it describes the late-time behavior of strings attached to heavy quark sources on the boundary. 

Having taken the ansatz, we can write down the induced metric in a form suitable for our further purposes. It is simply 

\begin{equation}\label{imetric}
ds^2=w R^2\Bigl(-\bigl(f-v^2\bigr)dt^2+2v\partial_r\xi\,dtdr+
\bigl(f^{-1}+(\partial_r\xi)^2\bigr)dr^2\Bigr)
\,,
\end{equation}
where $\partial_r\xi$ means $\frac{\partial\xi}{\partial r}$. Importantly, the induced metric has a horizon at $r=\rv$ because the equation $f(r)=v^2$ has a solution on the interval $[0,\rh ]$. We call it an induced horizon.\footnote{This naming convention is shorthand for saying that it is related to the induced metric on the worldsheet.} The Nambu-Goto action now takes the form

\begin{equation}\label{S12-1}
S=-\g\int dt\int_0^{\rb} dr \,w\sqrt{1-\frac{v^2}{f}+f(\partial_r\xi)^2}
\,,
\end{equation}
with $\g=\frac{R^2}{2\pi\alpha'}$.

In the presence of external forces exerted on the string endpoints, the total action $\cal S$ includes, in addition to the standard Nambu-Goto action, boundary terms arising from the coupling with external fields. For purposes of this paper, we do not need to make them explicit. What we do need is a variation of $\cal S$ with respect to the field $\xi$ as well as with respect to $x_a$, $x_b$ and $r_b$ describing the location of the string endpoints. Since the motion is uniform, we can choose $t=0$, as illustrated in Figure \ref{conf1-3}. Then $\delta\xi(0) = \delta x_a$ and $\delta\xi(r_b)=\delta x_b-\partial_r\xi\delta r_b$.\footnote{The latter follows from the chain rule because the field $\xi$ depends on $r_b$ according to Eq.\eqref{bcond}.} After a simple calculation, we find\footnote{This equation also defines our sign conventions for the $\sigma_i$'s and $F_i$'s.}

\begin{equation}\label{S12-2}
\delta {\cal S}=-\int dt\int_0^{\rb}dr\, \partial_r\sigma_x \,\delta\xi+\int dt\,
(\sigma_x-F_x)(\delta\xb-\delta\xa )
+(\sigma_r-F_r)\delta\rb
\,,
\end{equation}
with 
\begin{equation}\label{S12-3}
\sigma_x=-\frac{\g wf\partial_r\xi}{\sqrt{1-\frac{v^2}{f}+f(\partial_r\xi)^2}}
\,,
\qquad
\sigma_r=-\frac{\g w (1-\frac{v^2}{f})}{\sqrt{1-\frac{v^2}{f}+f(\partial_r\xi)^2}}
\,.
 \end{equation}
Here we assumed that the endpoint $A$ is fixed in the $r$-direction, and as a consequence of that, $\delta r_a=0$.

A pair of comments about equations \eqref{S12-2} and \eqref{S12-3} is in order. First, $\sigma_x=c$ is the first integral of the equation of motion for $\xi$ and therefore an absolute value of $\sigma_x$ must be the same at both endpoints. The force balance is the reason why $F_x$ must also obey this requirement. This all together determines the form of the boundary terms in \eqref{S12-2}. Second, the $\sigma_i$'s are nothing else but the components of the world-sheet current ${\cal P}^\sigma_\mu$ \cite{BZ}. Explicitly, $\sigma_x={\cal P}_x^\sigma=\frac{\delta S}{\delta\xi'}$ and 
$\sigma_r={\cal P}_r^\sigma=\frac{\delta S}{\delta r'}$, where a prime denotes a derivative with respect to the worldsheet coordinate $\sigma$. 

For what follows, it is convenient to express the integration constant $c$ in terms of $r_b$ and $\tan\a=\partial_r\xi\vert_{\rb}$, and perform the following rescaling 

\begin{equation}\label{theta}
\tan\a=\frac{\sqrt{1-\frac{v^2}{f}}}{\sqrt{f}}\,\tan\theta
\,.
\end{equation}
At the point $B$ the $\sigma$'s are then

\begin{equation}\label{sigma-tan}
\sigma_x=-\se\sin\theta
\,,
\qquad
\sigma_r=-\g w\sqrt{1-\frac{v^2}{f}}\cos\theta
\,,
\end{equation}
where $\se=\g w\sqrt{f}$. Note that $\sigma_x>0$ for configuration 1 and $\sigma_x<0$ for configuration 2.

The string length along the $x$-axis can be found by further integrating the 
first integral. First, we get from \eqref{S12-3}

\begin{equation}\label{xi}
\partial_r\xi=\pm\frac{1}{\sqrt{f}}\sqrt{\frac{1-\frac{v^2}{f}}{\frac{\se^2}{\sigma_x^2}-1}}
\,,
\end{equation}
with the minus sign for configuration 1 and the plus sign for 2. Then the integration over $r$ yields 

\begin{equation}\label{l12}
\ell=\vert\xa-\xb\vert=\rh\int^\nu_0\frac{d\rho 
}{f(\rh\rho)}
\sqrt{f(\rh\rho)-v^2}
\biggl(\frac{1}{\sin^2\theta}\,\frac{\se^2(\rh\rho)}{\se^2(\rh\nu)}-1\biggr)^{-\oh}
\,,
\end{equation}
where $\nu=\frac{\rb}{\rh}$. In this derivation we set $t=0$. It is clear that a similar derivation for $t\not =0$ would give the same result. 

We conclude our discussion of configurations 1 and 2 with the question of mechanical equilibrium at the endpoints. Clearly, a net force must vanish at each endpoint that translated into a mathematical language means precisely that the boundary terms vanish in \eqref{S12-2}. Thus we have 

\begin{equation}\label{balance12-0}
F_x-\sigma_x=0
\,,
\end{equation}
at the point $A$ and 

\begin{equation}\label{balance12-r0}
\sigma_x-F_x=0
\,,\qquad
\sigma_r-F_r=0
\,,
\end{equation}
at the point $B$. 

Now we will briefly describe the remaining configuration of Figure \ref{conf1-3} which is a semi-infinite string trailing out behind the point $B$. It is an extension of what was proposed for the $\text{AdS}$ geometry to that of \eqref{metric}. So, most of material comes from \cite{drag-ads}.

As before, we choose the static gauge $t=\tau$ and $r(\sigma)=c_1\sigma+c_0$ which, when combined with the boundary conditions

\begin{equation}\label{bcon-B}
r(0)=\rb\,,\qquad r(1)=\rh
\,,
\end{equation}
yields $c_0=r_b$ and $c_1=r_h-r_b$. For $x(t,r)$, we take the original ansatz \eqref{ansatz}. Then the Nambu-Goto action becomes 

\begin{equation}\label{S3-1}
S=-\g\int dt\int^{\rh}_{\rb} dr \,w\sqrt{1-\frac{v^2}{f}+f(\partial_r\xi)^2}
\,.
\end{equation}
In the presence of external forces exerted on the endpoint $B$, it should be modified to include boundary terms arising from the coupling with external fields. Again, we do not need to make them explicit. We need to know the variation of the total action with respect to the field $\xi$ and coordinates of $B$. Since the motion is uniform, we consider it at time $t=0$. Then as before, a little calculation shows that the variation of ${\cal S}$ is given by 

\begin{equation}\label{S3-2}
\delta{\cal S}=-\int dt\int_{\rb}^{\rh}dr\, \partial_r\sigma_x \,\delta\xi
+
\int dt\,(F_x-\sigma_x)\delta\xb+(F_r-\sigma_r)\delta\rb
\,,
\end{equation}
where $\sigma_x$ and $\sigma_r$ are given by \eqref{S12-3}. 

As one sees from the equation of motion, $\sigma_x=c$ is the first integral. The point is that the only way which allows one to avoid an imaginary right hand side in \eqref{xi} is to fix the integration constant at $r=\rv$. It is thus

\begin{equation}\label{I3}
\sigma_x=\sev 
\,,
\end{equation}
with $\sev=\se(r_v)$. From this, at the point $B$ one deduces 

\begin{equation}\label{sigmar3}
\sigma_r=-\g w\sqrt{\Bigl(1-\frac{v^2}{f}\Bigr)
\Bigl(1-\frac{\sev^2}{\se^2}\Bigr)}
\,.
\end{equation}
Note that, when written in this form, there is no manifest dependence on $\theta$.

Finally, we come to the question of mechanical equilibrium at the endpoint. The conditions for equilibrium can be read from the boundary terms in \eqref{S3-2}. We have

\begin{equation}\label{balance3}
F_x-\sigma_x=0\,,
\qquad
F_r-\sigma_r=0
\,.
\end{equation}

\subsubsection*{2. "Double" string configurations}

\begin{figure*}[htbp]
\centering
\includegraphics[width=4.95cm]{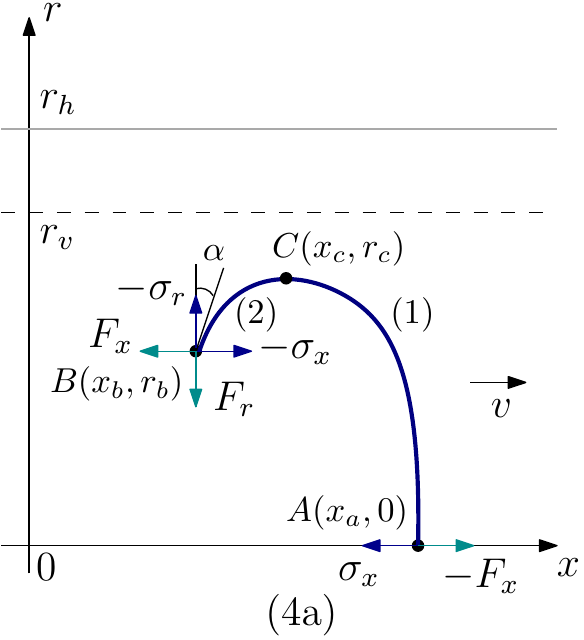}
\hspace{1.2cm}
\includegraphics[width=4.95cm]{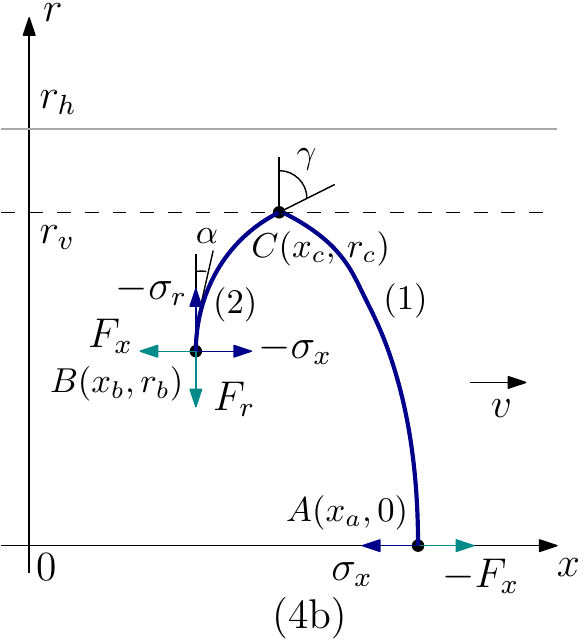}
\hspace{1.2cm}
\includegraphics[width=4.95cm]{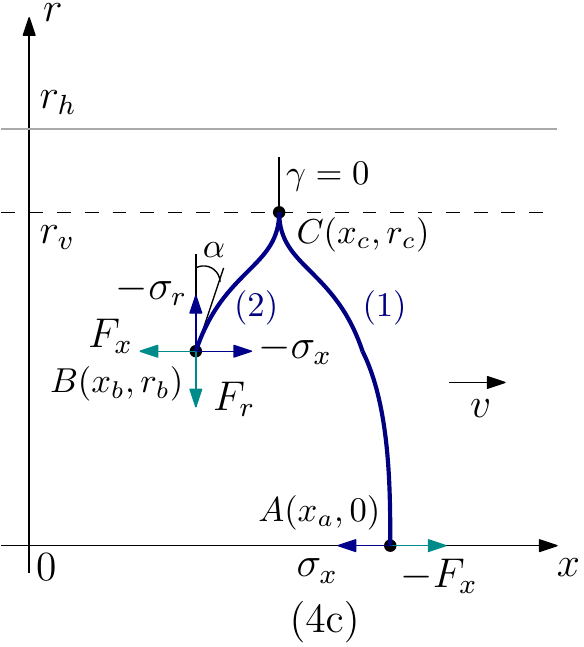}
\caption{{\small Double string configurations at time $t=0$. In all the cases the string moves with speed $v$ in the $x$-direction. $F_i$ stand for some external forces to balance the internal string tensions $\sigma_i$. In configurations 4b and 4c a cusp of angle $\gamma$ forms at the point $C$ lying on the induced horizon.}}
\label{conf4}
\end{figure*}

Having considered the configurations of Figure \ref{conf1-3}, we can now complete the description of our basis. To this end, we consider the configurations of Figure \ref{conf4}. 
There are two novelties involved. First, the approach we have used so far is now somewhat difficult to implement. The problem is that $\xi$ is a double valued function of $r$. A possible way out, which requires a little effort, is to think off a bending string as being glued from two elementary ones.\footnote{This is the reason why we call the configurations of Figure \ref{conf4} double string configurations.} Obviously, each string shown in Figure \ref{conf4} may be built by gluing together the endpoints of the two elementary strings $1$ and $2$ at the turning point $C$. The second, a cusp occurs on the binding string precisely at the point at which the string touches the induced horizon. The point is that if one considers a string bit near the point $C$, its speed approaches the local speed of light as the bit approaches the induced horizon. If so, the length of the bit in the lab frame contracts towards zero. In fact, this effect is general and not restricted to either the $\text{AdS}$ geometry \footnote{The occurrence of cusps on strings moving in $\text{AdS}_5$ is known in the literature \cite{pa}.} or its generalization given by \eqref{metric}. The main reason is the induced horizon. 

Before we start, let us briefly explain the appearance of these three configurations. Consider string $2$ which is stretched between the points $B$ and $C$. It is straightforward to repeat the previous analysis to show that the equation of motion can be integrated and then written as \eqref{xi}. If $\frac{\se^2}{\sigma_x^2}-1$ has a simple zero at $r=r_{\se}$ on the interval $[0,\rh]$, then there are three options: $r_{\se}<\rv$, $r_{\se}=\rv$, and $r_{\se}>\rv$. These precisely correspond to the configurations 4a, 4b, and 4c, shown in Figure \ref{conf4}.

We begin with configuration 4a. For the elementary strings 1 and 2, the variations of the corresponding actions can be read from \eqref{S12-2}. Dropping all but the boundary terms at the point $C$,  we have 

\begin{equation}\label{Si}
\delta S^{(i)}=\int dt\,\sigma_x^{(i)}\delta x_c+\sigma_r^{(i)}\delta r_c\,+\dots
\,,
\end{equation}
with $i=1,2$. Gluing the strings together requires that in the total action the boundary contributions cancel each other. So 

\begin{equation}\label{sewingC}
\sigma_x^{(1)}+\sigma_x^{(2)}=0
\,,\qquad
\sigma_r^{(1)}+\sigma_r^{(2)}=0
\,.
\end{equation}
It is easy to see that theses conditions are met at the turning point $C$, at which $r_c=r_{\se}$ holds. Indeed, we have $\sigma_x^{(i)}=\pm \se (r_c)$ and $\sigma_r^{(i)}=0$, since $\partial_r\xi^{(i)}\vert_{r_c}=\mp\infty$.\footnote{Here the upper sign refers to string 1 and the lower sign to string 2.} The latter follows directly from equation \eqref{xi}. 

Now consider string 2. It will be useful to write the $\sigma$'s at the point $B$ as

\begin{equation}\label{sigma4}
\sigma_x=\se\sin\theta
\,,
\qquad
\sigma_r=-\g w\sqrt{1-\frac{v^2}{f}}\,\cos\theta
\,,
\end{equation}
where we have rescaled $\tan\alpha=\partial_r\xi\vert_{r=\rb}$ so that 

\begin{equation}\label{theta2}
\tan\a=-\frac{\sqrt{1-\frac{v^2}{f}}}{\sqrt{f}}\,\tan\theta
\,.
\end{equation}
The odd-looking minus sign comes from keeping $\theta<0$ as it is for configuration 1 of Figure \ref{conf1-3}. Another useful relation is 

\begin{equation}\label{rm}
\se (r_c)=-\se\sin\theta
\,.
\end{equation}
It follows from the first integral of the equation of motion and relates $r_c$ to $r_b$ and $\theta$.

Given the expression \eqref{l12}, one can evaluate the string length along the $x$-direction, which is $\ell=\vert\xa-\xb\vert$, using the data at $B$ and gluing condition at $C$, with the result

\begin{equation}\label{l4}
\ell=\rh\int^\lambda_0\frac{d\rho }{f(\rh\rho)}
\sqrt{f(\rh\rho)-v^2}
\biggl(\frac{1}{\sin^2\theta}\,\frac{\se^2(\rh\rho)}{\se^2(\rh\nu)}-1\biggr)^{-\oh}
+
\rh\int^\lambda_\nu\frac{d\rho }{f(\rh\rho)}
\sqrt{f(\rh\rho)-v^2}
\biggl(\frac{1}{\sin^2\theta}\,\frac{\se^2(\rh\rho)}{\se^2(\rh\nu)}-1\biggr)^{-\oh}
\,.
\end{equation}
Here $\nu=\frac{\rb}{\rh}$ and $\lambda=\frac{r_c}{\rh}$. 

If one or more external forces are exerted on each string endpoints, then the conditions for mechanical equilibrium follow from \eqref{balance12-0} and \eqref{balance12-r0}. With the conventions shown in Figure \ref{conf4}, we get 

\begin{equation}\label{balance4a-A}
\sigma_x-F_x=0\,,
\end{equation}
at the point $A$ and 

\begin{equation}\label{balance4a-B}
F_x-\sigma_x=0\,,
\qquad
F_r-\sigma_r=0
\,,
\end{equation}
at the point $B$.

Configuration 4b may be analyzed along the above lines. The only novelty is that a cusp forms at the point $C$. This happens because $C$ lies on the induced horizon and the simple root is $r_{\se}=\rv$. The gluing conditions \eqref{sewingC} are satisfied by $\sigma_x^{(i)}=\pm \sev$ and $\sigma^{(i)}_r=0$. 

Now the angle $\theta$ at the endpoint $B$ is determined by 

\begin{equation}\label{thetaB}
\sin\theta=-\frac{\sev}{\se}
\,.
\end{equation}
As a result, the string length $\ell$ can be written as 

\begin{equation}\label{l4-1}
\ell=
\rh\int^{\frac{\rv}{\rh}}_0\frac{d\rho }{f(\rh\rho)}
\sqrt{f(\rh\rho)-v^2}
\biggl(\frac{\se^2(\rh\rho)}{\sev^2}-1\biggr)^{-\oh}
+
\rh\int^{\frac{\rv}{\rh}}_\nu\frac{d\rho }{f(\rh\rho)}
\sqrt{f(\rh\rho)-v^2}
\biggl(\frac{\se^2(\rh\rho)}{\sev^2}-1\biggr)^{-\oh}
\,.
\end{equation}
From the formula \eqref{xi}, it follows that the cusp (deviation) angle is given by 

\begin{equation}\label{cusp4b}
\tan\gamma=\frac{1}{v^2}\sqrt{\frac{\partial_rf}{2\partial_r\ln\se}}\,\,\bigg\vert_{\rv}
\,.
\end{equation}

Clearly, nothing special happens with the conditions for force balance at the string endpoints and, therefore, these are also given by \eqref{balance4a-A} and \eqref{balance4a-B}.

We conclude our discussion of the double string configurations with configuration 4c. In this case, we 
use the first integral of the equation of motion to parameterize the $\sigma_x$'s at the point $C$ as  
$\sigma_x^{(i)}=\pm\se\sin\theta (\rb)$  subject to the condition $\se\vert\sin\theta\vert(\rb)<\sev$. The latter guarantees a real right hand side in \eqref{xi}. As before, both $\sigma_r^{(i)}$ vanish at $C$.  All this together is enough to meet the gluing conditions at $C$.

Combining the first integral and equation \eqref{l12}, we find a useful formula for the string length along the $x$-direction

\begin{equation}\label{l4-2}
\ell=
\rh\int^{\frac{\rv}{\rh}}_0\frac{d\rho }{f(\rh\rho)}
\sqrt{f(\rh\rho)-v^2}
\biggl(\frac{1}{\sin^2\theta}\frac{\se^2(\rh\rho)}{\se^2(\rh\nu)}-1\biggr)^{-\oh}
+
\rh\int^{\frac{\rv}{\rh}}_\nu\frac{d\rho }{f(\rh\rho)}
\sqrt{f(\rh\rho)-v^2}
\biggl(\frac{1}{\sin^2\theta}\frac{\se^2(\rh\rho)}{\se^2(\rh\nu)}-1\biggr)^{-\oh}
\,.
\end{equation}
Since in the case of interest $r_{\se}>\rv$, we get from \eqref{xi} that the cusp angle is simply

\begin{equation}\label{cusp4c}
\tan\gamma=0
\,.
\end{equation}

Finally, there is no difficulty to understand that all the conditions for mechanical equilibrium are the same as before. 

\section{Gluing conditions}
\renewcommand{\theequation}{B.\arabic{equation}}
\setcounter{equation}{0}

The basic string configurations we have discussed in the Appendix A provide the blocks for building multi-string configurations. What is also needed are certain gluing conditions for strings meeting at baryon vertices (three-string junctions). Here we will describe such conditions. In doing so, we restrict ourself to the planar case so that all strings lie on the $xr$-plane.

\begin{figure*}[htbp]
\centering
\includegraphics[width=5cm]{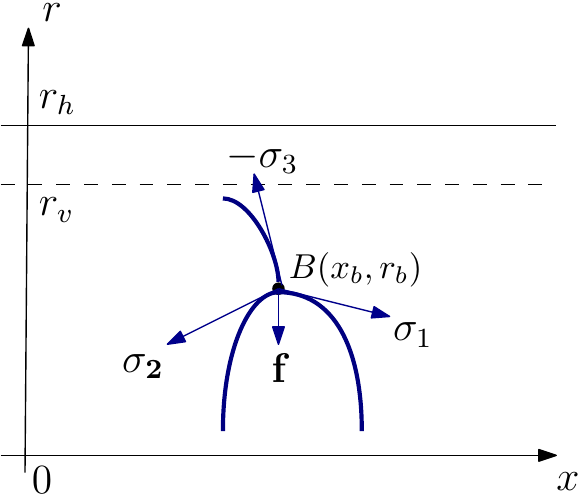}
\hspace{3cm}
\includegraphics[width=5cm]{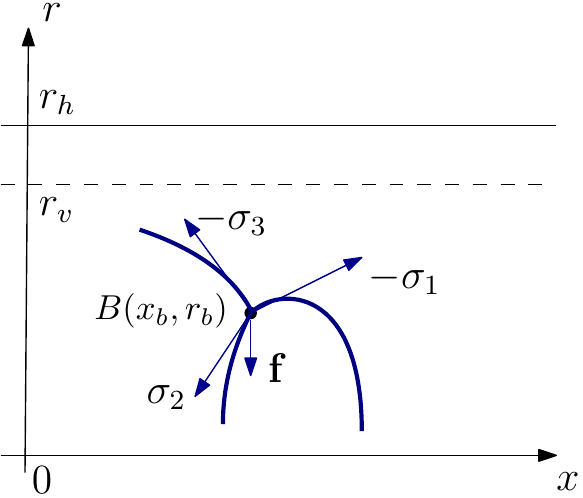}
\caption{{\small Three strings meeting in a baryon vertex placed at $B$. The gravitational force acting on the vertex is directed in the downward vertical direction. In vector notation, $\boldsymbol{\sigma}=(\sigma_x,\sigma_r)$ and $\mathbf{f}=(0,f_r)$. Left: The configuration is constructed from the configurations of Figure \ref{conf1-3}. Right: The configuration is constructed from the two configurations of Figure \ref{conf1-3} and one of Figure \ref{conf4}.}}
\label{balance-planar}
\end{figure*}

We begin with the configuration shown in Figure \ref{balance-planar}, on the left. The gluing conditions 
come from the requirement that in the variation of the total action with respect to a location of the vertex all boundary contributions cancel out. This is equivalent to a force balance condition at 
that vertex. Dropping all but the boundary terms at $B$, we find that the variation of the total action is 

\begin{equation}\label{B1}
\delta {\cal S}=\int dt\,
\bigl(\sigma_x^{(1)}+\sigma_x^{(2)}-\sigma_x^{(3)}\bigr)\delta\xb
+\bigl(\sigma_r^{(1)}+\sigma_r^{(2)}-\sigma_r^{(3)}+f_r\bigr)\delta\rb
\,\,+\dots
\,,
\end{equation}
where the $\sigma$'s can be read from the corresponding formulas of the Appendix A. The gravitation force acting on the vertex is 

\begin{equation}\label{force-vertex}
f_r=-m\,\partial_r\bigl(\sqrt{f-v^2}\,{\cal V}\bigr)
\,,
\end{equation}
as follows from \eqref{vertex-v}. After a little algebra, we find 

\begin{equation}\label{fb-x}
\sin\theta_1+\sin\theta_2+\phi=0
\,,
\end{equation}
and 
\begin{equation}\label{fb-r}
\cos\theta_1
+ \cos\theta_2
-
\sqrt{1-\phi^2}
+
3\kappa\frac{\sqrt{f}}{w}
\Bigl(\partial_r{\cal V}+\oh{\cal V}\partial_r\ln\bigl(f-v^2\bigr)\Bigr)
=0
\,,
\end{equation}
with $\phi=\frac{\sev}{\se}$ and $\kappa=\frac{m}{3\g}$. These conditions determine $\theta_1$ and $\theta_2$ as functions of $\rb$, $v$, and $\rh$. Note that $\theta_1$ is negative, while $\theta_2$ positive. $\phi$ takes values on the interval $[0,1]$ because $\se$ decreases with increasing $r$ and $r_b<\rv$.

It is straightforward to repeat the above analysis for the configuration shown in Figure \ref{balance-planar}, on the right. The only difference is that string 1 changes its shape from configuration 1 of Figure \ref{conf1-3} to one of those shown in Figure \ref{conf4}. In accordance with our conventions 
of the Appendix A, the condition \eqref{fb-x} holds but \eqref{fb-r} now becomes

\begin{equation}\label{fb-r2}
\cos\theta_1
-
\cos\theta_2
+
\sqrt{1-\phi^2}
-
3\kappa\frac{\sqrt{f}}{w}
\Bigl(\partial_r{\cal V}+\oh {\cal V}\partial_r\ln\bigl(f-v^2\bigr)\Bigr)
=0
\,.
\end{equation}
As before, these conditions determine the $\theta$'s in terms of $\rb$, $v$, and $\rh$. It is worth noting that if string 2 changes its shape to configuration 2 of Figure \ref{conf1-3}, then nothing happens with the form of the gluing conditions.

\section{Small $\nu$ expansion}
\renewcommand{\theequation}{C.\arabic{equation}}
\setcounter{equation}{0}

Here we explore in more detail the example discussed in Section III. Our goal is to compute the first two Taylor coefficients of $\ell$ expanded about $\nu=0$. The other coefficients can be computed in a similar way.

As already noted in subsection 4, it follows that for small $\nu$, only configuration I needs to be retained. If so, then $\ell$ is given by equation \eqref{lI} with the $\theta$'s determined from equations \eqref{fxI} and \eqref{frI}. It is quite natural to expand 

\begin{equation}\label{series}
\theta_i(\nu;v,h)=\sum_{n=0,\,n\,\text{even}}^\infty \theta_i^{(n)}\nu^n
\,,
\end{equation}
and then to plug these series into the gluing equations. After some algebra, we arrive at

\begin{equation}\label{thetas-nu}
\theta_i^{(0)}=\mp\sqrt{1-\varkappa^2}
\,,\qquad
\theta_i^{(2)}=\pm\frac{3}{2}\frac{\kappa\varkappa h}{\sqrt{1-\varkappa^2}}-\oh\frac{v}{\sqrt{1-v^2}}\ep^{h\sqrt{1-v^2}}
\,,
\end{equation}
where $\varkappa=\oh(1+3\kappa)$. The upper sign refers to $\theta_1$ and the lower sign to $\theta_2$.

Now plugging \eqref{thetas-nu} into \eqref{lI} and performing the integral over $\rho$, we get finally

\begin{equation}\label{l-nu}
\ell(\nu;v,h)=\frac{1}{\sqrt{\s}}\sqrt{1-v^2}\,\sqrt{h}\nu
\Bigl(\ell^{(1)}+\ell^{(3)}\,h\nu^2+O(\nu^4)\Bigr)
\,,
\end{equation}
with
\begin{equation}\label{l-nu2}
\ell^{(1)}=\frac{2}{3}\sqrt{1-\varkappa^2}\,{}_2F_1\bigl[\tfrac{1}{2},\tfrac{3}{4},\tfrac{7}{4},1-\varkappa^2\bigr]
\,,
\end{equation}

\begin{equation}\label{l-nu3}
\ell^{(3)}=
\frac{2+\varkappa-4\varkappa^2}{6\varkappa^2\sqrt{1-\varkappa^2}}
\biggl(5\,{}_2F_1\bigl[-\tfrac{1}{2},\tfrac{3}{4},\tfrac{7}{4},1-\varkappa^2\bigr]
-(2+\varkappa^2)\,{}_2F_1\bigl[\tfrac{1}{2},\tfrac{3}{4},\tfrac{7}{4},1-\varkappa^2\bigr]\biggr)
-\frac{\varkappa}{(1-\varkappa^2)^{\frac{3}{2}}}
\biggl(\frac{1}{\varkappa^2}-2+{}_2F_1\bigl[-\tfrac{1}{4},1,\tfrac{1}{4},1-\varkappa^2\bigr]\biggr)
\,.
\end{equation}
Here ${}_2F_1$ is the hypergeometric function. 

One point is worthy of note. To second order in $\nu$, the length transforms according to the Lorentz contraction formula. This is a puzzling result because there is no Lorentz invariance in the thermal medium which defines a preferred rest frame. The resolution is that such a transformation formula is violated by higher order terms.


\section{Taylor series of $\sigma_s$}
\renewcommand{\theequation}{D.\arabic{equation}}
\setcounter{equation}{0}

In Section IV, we needed to know the coefficients of the Taylor series of $\sigma_s(T,\mu)$ about the point $(T,0)$. Now we want to show that in fact these can be computed analytically. Here, as above, we restrict attention to the first two coefficients. 

Following \cite{a-screen}, we first deduce from the relations \eqref{tm} that 

\begin{equation}\label{h}
8xU=9
h^{\frac{3}{2}}\bigl(1-Zy^2\bigr)
\,,
\end{equation}
where  
\begin{equation}
U=\ep^{\frac{3}{2}h}-1-\frac{3}{2}h
\,,\qquad
Z=\frac{9+(7+6h)\ep^{-2h}-16\ep^{-\oh h}}{1-2\ep^{-\oh h}+\ep^{-h}}
\,.
\end{equation}
Here we have rescaled $T$ and $\mu$ as $T=\tfrac{\sqrt{\s}}{\pi}x$ and $\mu=2\mathfrak{r}\sqrt{3\s}\,y$. Expanding $h$ about $y=0$

\begin{equation}\label{h-exp}
h(x,y)=h_0(x)+h_2(x)y^2+O(y^4)
\,
\end{equation}
and then inserting it into \eqref{h}, we find 

\begin{equation}\label{h0h2}
x=\frac{9}{8}\h0^{\frac{3}{2}}U^{-1}
\,,
\qquad
h_2=\frac{2}{3}
\frac{\h0 ZU}
{U+\h0(1-\ep^{\frac{3}{2}\h0})}
\,\,.
\end{equation}
The first equation gives us $x$ as a function of $h_0$, while the second represents a recursion relation between $h_0$ and $h_2$. Note that from \eqref{h} it follows easily that the series \eqref{h-exp} contains only even powers of $y$. Therefore, \eqref{sigma-mu} is an expansion in even powers of $\mu$.

Now it is straightforward to compute the desired coefficients. Using the expansion for $h$ and the formula for the spatial string tension \eqref{drag-c}, after a short computation we get

\begin{equation}\label{coeffs}
\sigma^{(0)}_s=\sigma\frac{\ep^{\h0-1}}{\h0}
\,,\qquad
\sigma^{(2)}_s=
\frac{9\h0^3(\h0-1)Z\sigma_s^{(0)}}{128\pi\mathfrak{r}^2 U(U+\h0(1-\ep^{\frac{3}{2}\h0}))}
\,,
\end{equation}
with $\sigma=\g\s\ep$. Thus, the coefficients are given parametrically by equations \eqref{h0h2} and \eqref{coeffs}, where $h_0$ is a parameter. The other coefficients can be obtained similarly, but with more effort.


\end{document}